\documentclass[letterpaper,aps,prd,twocolumn,showpacs,superscriptaddress,groupedaddress,preprintnumbers,floatfix]{revtex4-1}  
\usepackage{graphicx}  
\usepackage{dcolumn}   
\usepackage{bm}        
\usepackage{amssymb}   

\usepackage{subfigure}
\usepackage{multirow}
\usepackage{slashbox}
\usepackage{placeins}
\usepackage{comment}

\usepackage{amsfonts}    
\usepackage{amsmath}

\begin{document}

\preprint{
\vbox{
\hbox{ADP-14-2/T860}
\hbox{Edinburgh 2014/01}
\hbox{LTH 1002}
\hbox{DESY 14-002}
}}

\widetext

\title{Magnetic form factors of the octet baryons \\from lattice QCD and chiral extrapolation}

\author{P.E.~Shanahan}\affiliation{ARC Centre of Excellence in Particle Physics at the Terascale and CSSM, School of Chemistry and Physics,
  University of Adelaide, Adelaide SA 5005, Australia}
\author{R.~Horsley}\affiliation{School of Physics and Astronomy, University of Edinburgh, Edinburgh EH9 3JZ, UK}
\author{Y.~Nakamura}\affiliation{RIKEN Advanced Institute for Computational Science, Kobe, Hyogo 650-0047, Japan}
\author{D.~Pleiter}\affiliation{JSC, Forschungzentrum J\"ulich, 52425 J\"ulich, Germany} \affiliation{Institut f\"ur Theoretische Physik, Universit\"at Regensburg, 93040 Regensburg, Germany}
\author{P.E.L.~Rakow}\affiliation{Theoretical Physics Division, Department of Mathematical Sciences, University of Liverpool, Liverpool L69 3BX, UK}
\author{G.~Schierholz}\affiliation{Deutsches Elektronen-Synchrotron DESY, 22603 Hamburg, Germany}
\author{H.~St\"uben}\affiliation{Regionales Rechenzentrum, Universit\"at Hamburg, 20146 Hamburg, Germany}
\author{A.W.~Thomas}\affiliation{ARC Centre of Excellence in Particle Physics at the Terascale and CSSM, School of Chemistry and Physics,
  University of Adelaide, Adelaide SA 5005, Australia}
\author{R.D.~Young}\affiliation{ARC Centre of Excellence in Particle Physics at the Terascale and CSSM, School of Chemistry and Physics,
  University of Adelaide, Adelaide SA 5005, Australia}
\author{J.M.~Zanotti}\affiliation{ARC Centre of Excellence in Particle Physics at the Terascale and CSSM, School of Chemistry and Physics,
  University of Adelaide, Adelaide SA 5005, Australia}

\collaboration{CSSM and QCDSF/UKQCD Collaborations}

\begin{abstract}
We present a 2+1--flavor lattice QCD calculation of the electromagnetic Dirac and Pauli form factors of the octet baryons. The magnetic Sachs form factor is extrapolated at six fixed values of $Q^2$ to the physical pseudoscalar masses and infinite volume using a formulation based on heavy-baryon chiral perturbation theory with finite-range regularization.
We properly account for omitted disconnected quark contractions using a partially-quenched effective field theory formalism.
The results compare well with the experimental form factors of the nucleon and the magnetic moments of the octet baryons. 
\end{abstract}


\maketitle

\section{Introduction}
\label{sec:Introduction}

The ability to accurately reproduce the experimentally determined baryon electromagnetic form factors stands as a fundamental test for any theoretical description of baryon structure. This is a test which quantum chromodynamics (QCD), our theory of the strong interactions, has not yet definitively passed~\cite{Arrington:2006zm}.

In particular, with ever-improving experimental determinations of the form factors revealing slight deviations from the phenomenological dipole form~\cite{Bernauer:2010wm,Jones:1999rz,Ron:2011rd,Zhan:2011ji}, a precise determination of these and related quantities from first-principles QCD is essential. Lattice simulation remains the only rigorous method to quantitatively probe the non-perturbative domain of QCD.
As well as facilitating a comparison of theory with experimental data, lattice simulations provide a great deal of physical insight and valuable information for model building by revealing the dependence of hadron properties on quark mass~\cite{Cloet:2002eg,Thomas:2002sj,Shanahan:2012wh}.

Recent years have seen a progression from quenched to fully dynamical lattice QCD studies of the electromagnetic form factors. Despite this significant advance, operator self-contractions (disconnected quark diagrams) are still often omitted from simulations as they are notoriously noisy and expensive to calculate. While in general this omission produces a systematic effect (shown to be small in Ref.~\cite{Abdel-Rehim:2013wlz}), exact results may be obtained for isovector quantities, where contributions from disconnected loops cancel.

We present new dynamical $2+1-$flavor lattice QCD simulation results for the electromagnetic form factors of the octet baryons. This data set includes results for $G_{E/M}$ for all outer-ring octet baryons at a range of discrete $Q^2$ values up to 1.3~GeV$^2$. 
As chiral extrapolations are different for the electric and magnetic form factors, we present here an analysis of the magnetic form factor only. $G_E$ will be considered in future work.

We extrapolate the lattice results for $G_M$, at each value of $Q^2$, as a function of quark mass to the physical point.
As the lattice simulations neglect disconnected quark contractions, this extrapolation is performed using a variation of partially-quenched chiral perturbation theory. The distinguishing feature of this formalism is that valence and sea quarks are treated separately. For example, one may set the electric charge of the sea quarks to zero, removing the same disconnected quark contractions omitted in the lattice simulations~\cite{Tiburzi:2009yd,Arndt:2003vd,Leinweber:2002qb}. This is termed `connected chiral perturbation theory'.
Finite-volume effects are estimated by using the leading one-loop results of the chiral effective field theory. 

By carrying out the lattice simulations over a range of light and strange quark masses it is possible to tightly constrain the chiral extrapolation on the relevant parameter space and obtain surprisingly accurate results for the form factors at the physical point. Those results compare quite favourably with the experimental values.

The details of the lattice simulation are given in Sec.~\ref{sec:LatticeSimulation}, while Sec.~\ref{sec:ChiPTextrap} presents the effective field theory formalism. Fits to the lattice simulation results are described in Sec.~\ref{sec:Fits}, followed by results for the magnetic isovector form factors, octet baryon magnetic moments and magnetic radii in Sec.~\ref{sec:Results}. The appendices provide further details, including tables of lattice results and functional forms for the chiral expansions.

\section{Lattice simulation}
\label{sec:LatticeSimulation}

\begin{table}[]
\caption{\label{tab:SimDetails} Simulation details for the ensembles used here, with $\beta=5.50$ corresponding to $a=0.074(2)$~fm. The scale is set using various singlet quantities~\cite{Horsley:2013wqa,Bietenholz:2011qq,Bietenholz2010436}. $L^3\times T=32^3 \times 64$ for all ensembles. The parameter $\kappa_0$ denotes the value of $\kappa_l=\kappa_s$ at the SU(3) symmetric point.}
\begin{ruledtabular}
\begin{tabular}{ccccccc}
 & $\kappa_0$ & $\kappa_l$ & $\kappa_s$  & $m_\pi$~(MeV) & $m_K$~(MeV) & $m_\pi L$ \\
\hline
1 & 0.120900 & 0.120900 & 0.120900 &   465 & 465 & 5.6 \\
2 &  & 0.121040 & 0.120620 & 360 & 505 & 4.3 \\
3 &  & 0.121095 & 0.120512 & 310 & 520 & 3.7 \\ \hline
4 & 0.120920 & 0.120920 & 0.120920 & 440 & 440 & 5.3 \\ \hline
5 & 0.120950 & 0.120950 & 0.120950 & 400 & 400 & 4.8 \\
6 &  & 0.121040 & 0.120770 & 330 & 435 & 4.0 
\end{tabular}
\end{ruledtabular}
\end{table}

Here we describe our lattice setup and summarize the standard methods used to calculate the octet baryon electromagnetic form factors. While the nucleon form factors have been investigated in many lattice studies~\cite{Alexandrou:2007xj,Alexandrou:2006ru,Gockeler:2003ay,Hagler:2007xi,Lin:2008uz,Liu:1994dr,Sasaki:2007gw,Hagler:2009ni,Boinepalli:2006xd,Collins:2011mk,Bratt:2010jn,Syritsyn:2009mx}, we emphasize that the results presented here also include values for the hyperon form factors, which have so far received only limited attention~\cite{Boinepalli:2006xd,Lin:2008mr,Wang:2008vb,Leinweber:1990dv}.
These are of significant interest both in their own right and because they provide valuable insight into the environmental sensitivity of the distribution of quarks inside a hadron. For example, one may learn how the distribution of $u$ quarks in the proton differs from that in the $\Sigma^+$, an effect caused by the mass difference of the spectator $d$ and $s$ quarks.

\subsection{Simulation parameters}

\begin{figure}[]
\begin{center}
\includegraphics[width=0.4\textwidth]{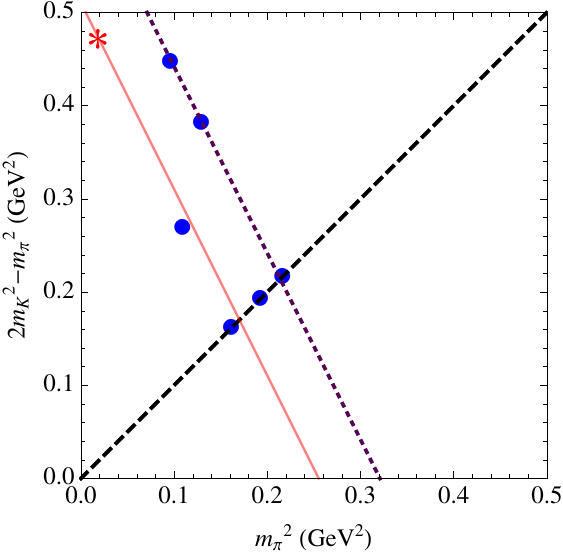}
\caption{Locations of lattice simulation results in the $m_l-m_s$ plane. The red star denotes the physical point and the dashes indicate the flavor-symmetric line where $m_l=m_s$. Our primary simulation trajectory, illustrated by the dotted line, corresponds to the line of constant singlet quark mass $(2m_K^2+m_\pi^2)$ at $\kappa_0=0.120900$ (simulations 1--3 in Table~\ref{tab:SimDetails}). The solid line indicates the physical value of the singlet mass.}
\label{fig:mlmsPlot}
\end{center}
\end{figure}

We use gauge field configurations with $N_f=2+1$ flavors of non-perturbatively $\mathcal{O}(a)$-improved Wilson fermions. The clover action consists of the tree-level Symanzik improved gluon action together with a mild `stout' smeared fermion action~\cite{Bietenholz2010436}. As the main aim of the work presented here is to perform a chiral (as well as infinite volume) extrapolation of the baryon electromagnetic form factors at fixed values of $Q^2$, we restrict ourselves to a single lattice volume of $L^3\times T=32^3 \times 64$. The lattice scale $a=0.074(2)$~fm is set using various singlet quantities~\cite{Horsley:2013wqa,Bietenholz:2011qq,Bietenholz2010436}. The lightest pion mass is about 310~MeV.
A summary of the simulation parameters is given in Table~\ref{tab:SimDetails}.

A particular feature of the gauge configurations is that the primary simulation trajectory in quark-mass space, illustrated in Fig.~\ref{fig:mlmsPlot}, follows a line of constant singlet mass $m_q = (m_u +m_d +m_s)/3 = (2m_l +m_s)/3$. This is achieved by first finding the SU(3) flavor-symmetric point where flavor singlet quantities take on their physical values, then varying the individual quark masses about that point~\cite{Bietenholz:2011qq,Bietenholz2010436}. 

It is clear from Fig.~\ref{fig:mlmsPlot} that this primary trajectory at $\kappa_0=0.120900$ (where $\kappa_0$ denotes the value of $\kappa_l=\kappa_s$ at the SU(3) symmetric point) does not quite match the physical singlet mass line~\cite{Bietenholz:2011qq}. 
Extrapolation to the physical point thus requires a shift not only along the simulation trajectory but in a direction perpendicular to it. To constrain the quark-mass dependence in this perpendicular direction we include additional lattice simulations at several singlet masses (i.e., values of $\kappa_0$). These are listed as simulations $4-6$ in Table~\ref{tab:SimDetails} and are shown in Fig.~\ref{fig:mlmsPlot}.

\subsection{Electromagnetic form factors}

The Dirac and Pauli form factors $F_1(Q^2)$ and $F_2(Q^2)$ are obtained from the standard decomposition of the matrix elements of the electromagnetic current $j_\mu$:
\begin{align}
\nonumber
\langle B(p'&,s') | j_\mu(q) | B(p,s)\rangle = \\
&\overline{u}(p',s') \left[ \gamma_\mu F_1(Q^2)+  \frac{i\sigma_{\mu\nu}q^\nu}{2m_B}F_2(Q^2) \right]u(p,s),
\label{eq:FFdecomp}
\end{align}
where $u(p,s)$ is a Dirac spinor with momentum $p$ and spin polarization $s$, $q=p'-p$ is the momentum transfer, $Q^2=-q^2$ and $m_B$ is the mass of the baryon B.

The left hand side of Eq.~(\ref{eq:FFdecomp}) is calculated on the lattice in the usual way from the ratio of three- and two-point correlation functions:
\begin{align} \nonumber
R(t,\tau;\vec{p}\,',& \vec{p}) =  \frac{C_{3\textrm{pt}}(t,\tau;\vec{p}\,',\vec{p})}{C_{2\textrm{pt}}(t,\vec{p}\,')} \\
& \times \left[ \frac{C_{2\textrm{pt}}(\tau,\vec{p}\,')C_{2\textrm{pt}}(\tau,\vec{p}\,')C_{2\textrm{pt}}(t-\tau,\vec{p})}{C_{2\textrm{pt}}(\tau,\vec{p})C_{2\textrm{pt}}(t,\vec{p})C_{2\textrm{pt}}(t-\tau,\vec{p}\,')}\right]^{1/2},
\end{align}
where $t$ denotes the Euclidean time position of the sink and $\tau$ the operator insertion time. In order to ensure that excited state contributions to the correlation functions are suppressed, we employ quark smearing at the source and sink and use a generous source-sink separation of $1-1.15$~fm. This has been shown to be sufficient~\cite{Collins:2011mk}.

\begin{figure}
\begin{center}
\subfigure[Doubly-represented quark contributions.]{
\includegraphics[width=0.48\textwidth]{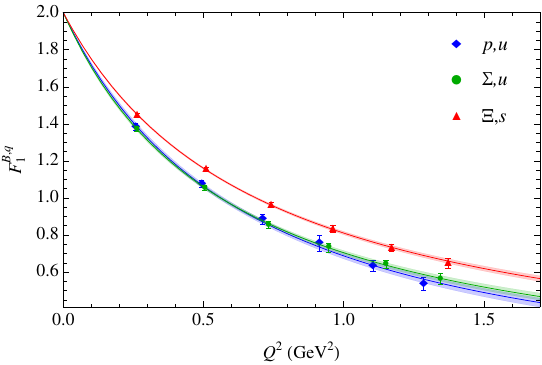}
\label{fig:F1doubly}
}
\subfigure[Singly-represented quark contributions.]{
\includegraphics[width=0.48\textwidth]{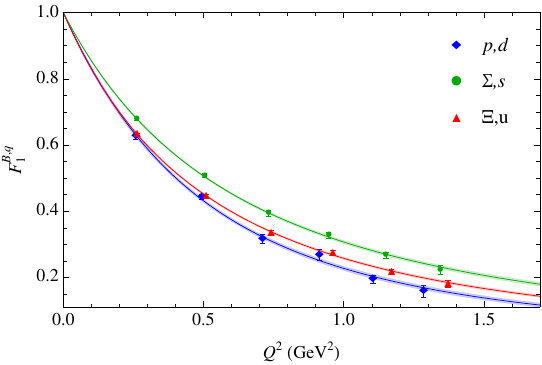}
\label{fig:F1singly}
}
\caption{Quark contributions to the Dirac form factor $F_1$ of the hyperons at the lightest simulation pion mass $(m_\pi,m_K)=(310,520)~$MeV. The charges of the relevant quarks have been set to unity. The lines show dipole-like fits (Eq.~(\ref{eq:F1Fit})).}
\label{fig:F1}
\end{center}
\end{figure}

The two- and three-point functions are given, as in Ref.~\cite{Collins:2011mk}, by
\begin{align}
C_{2pt}(\tau,\vec{p}) & = \textrm{Tr}\left[ \frac{1}{2}(1+\gamma_4) \langle{B}(\tau,\vec{p})\overline{B}(0,\vec{p})\rangle\right], \\
C_{3pt}(t,\tau,\vec{p}\,',\vec{p},\mathcal{O}) & = \textrm{Tr}\left[ \Gamma\langle{B}(t,\vec{p}\,')\mathcal{O}(\vec{q},\tau)\overline{B}(0,\vec{p})\rangle\right], \label{eq:3pt}
\end{align}
where `Tr' denotes a trace in spinor space and the local vector current $\mathcal{O}$ is
\begin{equation}
\mathcal{O}_\mu(\vec{q},\tau) = \sum_{\vec{x}}e^{i\vec{q}\cdot\vec{x}} \overline{q}(\vec{x},\tau)\gamma_\mu q(\vec{x},\tau), \label{eq:O}
\end{equation}
where $q(\vec{x},\tau)$ is a quark field and $\vec{q}$ is the three-momentum transfer.
The Dirac operator $\Gamma$ represents a polarization projection. For example, we use
\begin{align}
\Gamma_\textrm{unpol.} & = \frac{1}{2}(1+\gamma_4), \\
\Gamma_3 & = \frac{1}{2}(1+\gamma_4)i\gamma_5\gamma_3,
\end{align}
for an unpolarized baryon or one polarized in the $z$-direction respectively. As the current $\mathcal{O}$ is not conserved, we enforce charge conservation by using $2/F^{p,u}_1(0)$ as a multiplicative renormalization on each ensemble (as explained later, the quark-level form factors are defined for quarks of unit charge). We note that quark line disconnected contributions to the three-point function of Eq.~(\ref{eq:3pt}) are neglected in these simulations. The effect of this omission will be discussed further in the following sections.
Simulations are performed with zero sink momentum and six different values of the momentum transfer $\vec{q}=\vec{p}\,'-\vec{p}$, corresponding to $Q^2$ up to $\approx$1.3~GeV$^2$.

For the chiral extrapolations presented in this work we consider only linear combinations of $F_1$ and $F_2$, namely the electric and magnetic Sachs form factors:
\begin{align}
G_E(Q^2)&=F_1(Q^2)-\frac{Q^2}{4m_N^2}F_2(Q^2), \\G_M(Q^2)&=F_1(Q^2)+F_2(Q^2).
\end{align}
We focus particularly on the magnetic form factor $G_M$. 

\begin{figure}
\begin{center}
\subfigure[Doubly-represented quark contributions.]{
\includegraphics[width=0.48\textwidth]{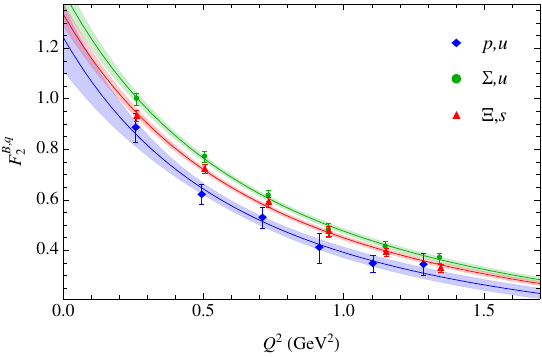}
\label{fig:F2doubly}
}
\subfigure[Singly-represented quark contributions.]{
\includegraphics[width=0.48\textwidth]{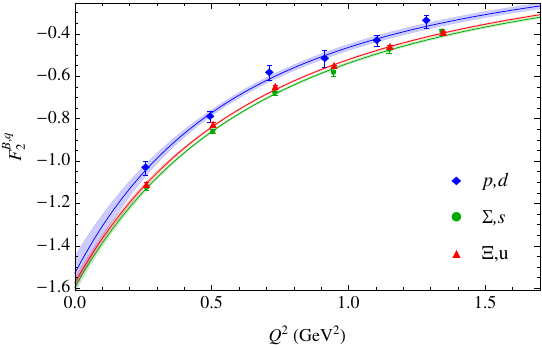}
\label{fig:F2singly}
}
\caption{Quark contributions to the Pauli form factor $F_2$ of the hyperons at the lightest simulation pion mass $(m_\pi,m_K)=(310,520)~$MeV. The charges of the relevant quarks have been set to unity. The lines show dipole fits (Eq.~(\ref{eq:F2Fit})).}
\label{fig:F2}
\end{center}
\end{figure}

\begin{figure}
\begin{center}
\subfigure[Isovector Dirac radii.]{
\includegraphics[width=0.48\textwidth]{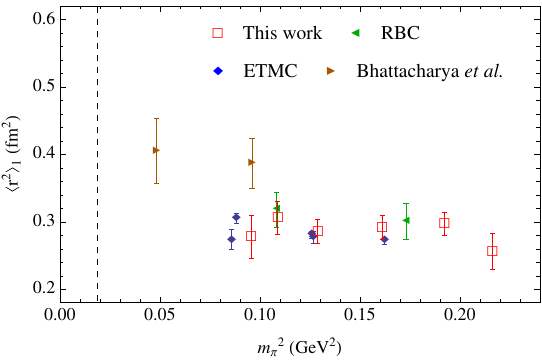}
\label{fig:r1}
}
\subfigure[Isovector Pauli radii.]{
\includegraphics[width=0.48\textwidth]{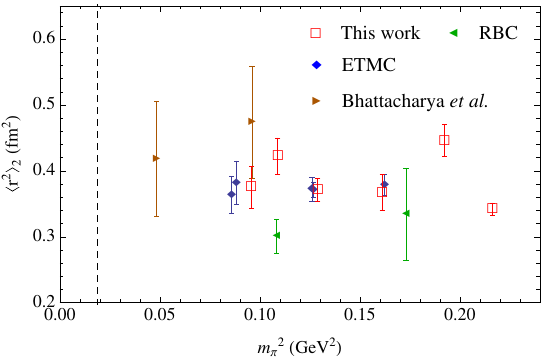}
\label{fig:r2}
}
\end{center}
\caption{Dirac and Pauli radii for the nucleon from recent $2+1$ and $2+1+1$-flavor lattice simulations~\cite{Bhattacharya:2013ehc,Bratt:2010jn,Lin:2010fv,Syritsyn:2009mx,Yamazaki:2009zq}, compared with the results of this work.}
\label{fig:Radii}
\end{figure}

\subsection{Lattice results for $F_1$ and $F_2$}
\label{subsec:F1F2res}

Although the primary focus of this work is the values of the magnetic form factors at the physical quark masses, with details of the chiral extrapolation of $G_M$ presented in the following sections, we display here some of the raw lattice simulation results for $F_{1,2}$ before finite-volume corrections or chiral extrapolations have been applied. Numerical results are tabulated in Appendix~\ref{app:rawLatticeResults}. We also give the results of a dipole-like extraction of the Dirac and Pauli mean-squared radii and the anomalous magnetic moment.

The Dirac and Pauli form factors at the lightest simulation pion mass $(m_\pi,m_K)=(310,520)~$MeV are illustrated in Figs.~\ref{fig:F1} and \ref{fig:F2}. The figures have been organized as doubly and singly represented quark contributions. This grouping shows most clearly the environmental sensitivity of the quark contributions to the form factors; for example, the only difference between the $u$ in the proton and the $u$ in the sigma baryon is the mass of the spectator ($d$ or $s$) quark. 
For $F_1$ this sensitivity increases with $Q^2$.
The fits shown use the 2-parameter ans\"atze: 
\begin{align}
\label{eq:F1Fit}
F_1(Q^2) & = \frac{F_1(0)}{1 + c_{12}Q^2 + c_{14}Q^4}, \\ \label{eq:F2Fit}
F_2(Q^2) & = \frac{F_2(0)}{(1 + c_{22}Q^2)^2},
\end{align}
where the $c_{ij}$ and the anomalous magnetic moment $F_2^{B,q}(0) =\kappa^{B,q}$ are fit parameters, while $F_1(0)$ is fixed by charge conservation. As we consider quarks of unit charge, $F_1(0)=2,1$ for the doubly and singly represented quarks respectively. Clearly, the functional forms chosen provide excellent fits to the lattice simulation results.  

Mean-squared radii are extracted from the $Q^2$-derivatives of the form factors:
\begin{equation}
\langle r^2 \rangle_i = -\frac{6}{F_i(0)} \frac{d}{dQ^2} F_i(Q^2) \bigg |_{Q^2=0}.
\end{equation}
The isovector radii for the nucleon are shown in Fig.~\ref{fig:Radii}. These results are in line with those from other $2+1$ and $2+1+1$--flavor simulations~\cite{Bhattacharya:2013ehc,Bratt:2010jn,Lin:2010fv,Syritsyn:2009mx,Yamazaki:2009zq}. We note that the other lattice simulations included here were performed at a range of values of $m_K$. Although most results were extracted using dipole or dipole-like fits, some include a systematic uncertainty arising from that choice of fit function while others do not. This partially accounts for the large variation in the quoted errors.
Tables of results for all $\langle r^2 \rangle_{1,2}^{B,q}$ and $\kappa^{B,q}$ extracted from our fits are given in Appendix~\ref{app:rawLatticeResults}.

\section{Connected chiral perturbation theory extrapolation}
\label{sec:ChiPTextrap}

While lattice QCD has made great progress towards a quantitative understanding of the physics of the strong interaction in the non-perturbative regime, it is often necessary to extrapolate lattice results from unphysically large simulation meson masses to the physical point. Partially-quenched chiral perturbation theory has been developed in order to address the extrapolation of partially-quenched lattice studies, which employ different values for the sea and valence quark masses.

The lattice simulations considered here, although fully dynamical, include only contributions from `connected' insertions of the current operator. For this reason, we extrapolate the results using a formalism based on `connected chiral perturbation theory', which is a variant of partially-quenched chiral perturbation theory.

Partially-quenched chiral perturbation theory has been widely discussed in the literature~\cite{Bernard:1993sv,Sharpe:2001fh,Sharpe:2000bc,Sharpe:1995qp,Chen:2001yi,Savage:2001dy,Leinweber:2002qb,Allton:2005vm}. Here we employ the heavy-baryon expansion pioneered by Jenkins and Manohar~\cite{Jenkins:1990jv,Jenkins:1991ts,Jenkins:1991es,Jenkins:1991bt,Jenkins:1992pi}. For completeness, this section summarizes our adaptation of this formalism and the relevant expressions for the magnetic form factors of the octet baryons.

\subsection{Partially-quenched chiral perturbation theory}

Details of partially-quenched chiral perturbation theory may be found in Refs.~\cite{Bernard:1993sv,Sharpe:2001fh,Sharpe:2000bc,Sharpe:1995qp,Chen:2001yi,Savage:2001dy,Leinweber:2002qb,Allton:2005vm}. Here we outline a special case of this formalism, termed `connected chiral perturbation theory'~\cite{Tiburzi:2009yd}.

Partially-quenched QCD includes nine quarks, which appear in the fundamental representation of the graded symmetry group $SU(6|3)$:
\begin{equation}
\psi^T = \left( u,d,s,j,l,r,\tilde{u},\tilde{d},\tilde{s} \right).
\end{equation}
In addition to the three usual light quarks $(u,d,s)$, there are three light fermionic sea quarks $(j,l,r)$ and three spin-1/2 bosonic ghost quarks $\left( \tilde{u},\tilde{d},\tilde{s}\right)$. When the ghost quarks are made pairwise mass- and charge-degenerate with $(u,d,s)$, their bosonic statistics ensure that closed $q$ and $\tilde{q}$ quark loop contributions cancel and hence such loops do not contribute to observables. Thus, if only $(u,d,s)$ are used in hadronic interpolating fields, these quarks truly represent `valence' quarks, while $\left(j,l,r\right)$ appear only in disconnected loops and are therefore interpreted as sea quarks.

For our application, the sea and ghost quarks are mass-degenerate with their corresponding valence partners. Thus, the quark mass matrix is
\begin{equation}
m_\psi = \textrm{diag}\left(m_u,m_d,m_s,m_u,m_d,m_s,m_u,m_d,m_s\right).
\end{equation}

As we wish to exclude all diagrams with closed quark loops from contributing to hadronic observables, we set the sea quark charges to zero.
As the ghost quarks $\left( \tilde{u},\tilde{d},\tilde{s}\right)$ must have the same charges, pairwise, as $(u,d,s)$,
the general form of the quark charge matrix is
\begin{equation}
Q=\textrm{diag}\left(q_u,q_d,q_s,0,0,0,q_u,q_d,q_s\right).
\end{equation}
Individual quark contributions may be extracted by setting all but one charge to zero, for example by taking
$q_u\rightarrow 1$, $q_d \rightarrow 0$, $q_s \rightarrow 0$ to obtain the $u$-quark contribution.
Of course, reinstating the sea quark charges by $Q \rightarrow \textrm{diag}\left(q_u,q_d,q_s,q_u,q_d,q_s,q_u,q_d,q_s\right)$ will give a formalism which reproduces exactly full chiral perturbation theory~\cite{Sharpe:1995qp}.

The dynamics of the 80 pseudo-Goldstone mesons (both bosonic and fermionic) which emerge from the spontaneous breaking of the symmetry group $SU(6|3)_L\otimes SU(6|3)_R\otimes U(1)_V \rightarrow SU(6|3)_V\otimes U(1)_V$ are described at lowest order by the Lagrangian
\begin{equation}
\label{eq:lambdaeq}
\mathcal{L}=\frac{f^2}{8}\textrm{Str}\left(D^\mu\Sigma^\dagger D_\mu\Sigma\right) + \lambda \textrm{Str}\left(m_\psi\Sigma+m_\psi^\dagger \Sigma^\dagger \right),
\end{equation}
where
\begin{equation}
\Phi = \left( \begin{array}{cc} M & \chi^\dagger \\ \chi & \widetilde{M} \end{array} \right ), \hspace{.4cm} \Sigma = \xi^2 = \textrm{exp}\left(\frac{2i\Phi}{f}\right).
\end{equation}
Here $M$, $\widetilde{M}$ and $\chi$ are matrices of pseudo-Goldstone bosons with the quantum numbers of $q\overline{q}$ pairs, pseudo-Goldstone bosons with the quantum numbers of $\tilde{q}\overline{\tilde{q}}$ pairs, and pseudo-Goldstone fermions with the quantum numbers of $\tilde{q}\,\overline{q}$ pairs, respectively. Made explicit in Ref.~\cite{Chen:2001yi}, $\Phi$ is normalized such that $\Phi_{12} =\pi^+$. $\textrm{Str}$ denotes the supertrace.
The gauge-covariant derivative is given by $D_\mu \Sigma = \partial_\mu \Sigma + ie \mathcal{A}_\mu \left[Q,\Sigma\right]$.

While the complete partially-quenched theory includes baryons composed of all types (and all mixtures of types) of quarks, for our application we need only predominantly valence states with at most one ghost or sea quark. These are constructed explicitly in Ref.~\cite{Chen:2001yi}. In general terms, the baryon field $B_{ijk}$ is constructed using an interpolating field
\begin{equation}
B_{ijk}^\gamma \sim \left( \psi_i^{\alpha,a}\psi_j^{\beta,b}\psi_k^{\gamma,c}-\psi_i^{\alpha,a}\psi_j^{\gamma,c}\psi_k^{\beta,b}\right)\epsilon_{abc}(C\gamma_5)_{\alpha\beta}.
\end{equation}
The usual spin-1/2 baryon octet is embedded in $B_{ijk}$ for $i,j,k$ restricted to 1--3 as
\begin{equation}
B_{ijk}=\frac{1}{\sqrt{6}}\left(\epsilon_{ijl}{\bf B}^l_k+\epsilon_{ikl} {\bf B}^l_j \right),
\end{equation}
where
\begin{equation}
{\bf B} = \left(\begin{array}{ccc} \frac{1}{\sqrt{6}}\Lambda + \frac{1}{\sqrt{2}}\Sigma^0 & \Sigma^+ & p \\\Sigma^- & \frac{1}{\sqrt{6}}\Lambda-\frac{1}{\sqrt{2}}\Sigma^0 & n\\
\Xi^- & \Xi^0 & -\frac{2}{\sqrt{6}}\Lambda\end{array}\right).
\end{equation}
Similarly, the spin-3/2 decuplet baryons may be constructed as
\begin{align}
\nonumber
T^{\alpha,\mu}_{ijk} \sim \left( \right.& \psi_i^{\alpha,a}\psi_j^{\beta,b}\psi_k^{\gamma,c} + \psi_i^{\beta,b}\psi_j^{\gamma,c}\psi_k^{\alpha,a} \\
& +\left. \psi_i^{\gamma,c}\psi_j^{\alpha,a}\psi_k^{\beta,b} \right)\epsilon_{abc}(C\gamma^\mu)_{\beta,\gamma},
\end{align}
where, for $i,j,k=$1--3, $T_{ijk}$ is simply the usual totally symmetric tensor containing the decuplet of valence baryon resonances.

The covariant derivative takes the same form for both the octet and decuplet baryons:
\begin{align}
\nonumber
\left(D^\mu B \right)_{ijk} = & \partial^\mu B_{ijk} + \left(V^\mu\right)_{li}B_{ljk}\\ \nonumber
&+(-1)^{\eta_i (\eta_j+\eta_m)}(V^\mu)_{jm}B_{imk}\\
&+(-1)^{(\eta_i+\eta_j)(\eta_k+\eta_n)}(V^\mu)_{kn}B_{ijn}.
\end{align}
Here the grading factor $\eta_k$ tracks the statistics of the bosonic ghost quark sector:
\begin{align}
\eta_k = 
\begin{cases}
1 & \textrm{for }k=1-6 \\
0 & \textrm{for }k=7-9,
\end{cases}
\end{align}
and the vector field $V^\mu$ is defined in analogy with that in QCD:
\begin{align}
V^\mu & = \frac{1}{2}\left(\xi \partial^\mu\xi^\dagger + \xi^\dagger\partial^\mu\xi\right).
\end{align}

The coupling of the 80 pseudo-Goldstone mesons to the baryons is described by
\begin{align}
\nonumber
\mathcal{L}= & 2\alpha \left( \overline{B} S^\mu B A_\mu \right) + 2\beta\left(\overline{B}S^\mu A_\mu B\right) \\ \nonumber
& + 2\gamma\left(\overline{B}S^\mu B\right) \textrm{Str}(A_\mu) + 2\mathcal{H}\left(\overline{T}^\nu S^\mu A_\mu T_\nu \right) \\ \nonumber
& + \sqrt{\frac{3}{2}}\mathcal{C} \left[ \left( \overline{T}^\nu A_\nu B \right) + \left( \overline{B} A_\nu T^\nu \right) \right] \\
& + 2\gamma^{'} \left(\overline{T}^\nu S^\mu T_\nu \right) \textrm{Str}(A_\mu),
\end{align}
where, again in analogy with QCD,
\begin{align}
A^\mu & = \frac{i}{2}\left(\xi \partial^\mu \xi^\dagger - \xi^\dagger \partial^\mu \xi\right),
\end{align}
$S^\mu$ is the covariant spin vector and the brackets are a shorthand for field bilinear invariants employed in Ref.~\cite{Labrenz:1996jy}, as summarized in Appendix~\ref{app:FBI}.
By matching to the usual QCD Lagrangian for $i,j,k$ restricted to 1--3, we make the identification
\begin{equation}
\alpha=\frac{2}{3}D+2F, \hspace{.2cm} \beta=-\frac{5}{3}D+F,
\end{equation}
while $\mathcal{C}$ and $\mathcal{H}$ map directly to their QCD values.

The heavy-baryon propagators for the octet baryon, decuplet baryon and meson are~\cite{Jenkins:1992pi}
\begin{equation}
\frac{i}{v\cdot k + i \epsilon}, \hspace{.2cm} \frac{i P^{\mu\nu}}{v\cdot k - \delta + i\epsilon}, \textrm{~~and~~} \frac{i}{k^2-M^2 + i\epsilon}
\end{equation}
respectively. Here $P^{\mu\nu}=v^\mu v^\nu - g^{\mu\nu} - (4/3)S^\mu S^\nu$ is a spin-polarization projector that projects out the positive spin-1/2 solutions to the equation of motion, and $\delta$ denotes the average octet-baryon--decuplet-baryon mass splitting.

\subsection{Electromagnetic form factors}

In the heavy-baryon formalism, the electromagnetic form factors $G_E$ and $G_M$ are defined by
\begin{align}
\label{eq:jstruct}
\langle B(p') | J_\mu | B(p) \rangle = \overline{u}(p') & \left\{  \vphantom{\frac{1}{2}} \right. v_ \mu G_E(Q^2)  \\ \nonumber
& + \left. \frac{i \epsilon_{\mu \nu \alpha \beta} v^\alpha S^\beta q^\nu}{m_N} G_M(Q^2) \right\} u(p),
\end{align}
where $q=p'-p$ and $Q^2=-q^2$. Here we focus exclusively on the magnetic form factor $G_M$, at fixed finite $Q^2$.

In the familiar formulation of chiral perturbation theory, the
magnetic moments of the octet baryons in the chiral limit are encoded in the coefficients of the `magnetic Lagrangian density'~\cite{Jenkins:1992pi}:
\begin{align}
\nonumber
\mathcal{L}= \frac{e}{4m_N}F_{\mu\nu}\sigma^{\mu\nu} \left[ \vphantom{\left(\overline{B}\right)}\right.& \mu_\alpha \left( \overline{B} B Q \right) + \mu_\beta \left(\overline{B} Q B \right)  \\
& \left. +\mu_\gamma\left(\overline{B}B\right)\textrm{Str}(Q) \right].
\label{eg:MagM}
\end{align}
By comparison with the standard QCD Lagrangian, we make the identification 
\begin{equation}
\label{eq:muDF}
\mu_\alpha = \frac{2}{3}\mu_D+2\mu_F, \hspace{3mm} \mu_\beta=-\frac{5}{3}\mu_D+\mu_F.
\end{equation}
The $\mu_\gamma$ term contributes only when the quark charge matrix $Q$ is defined such that $\textrm{Str}(Q) \ne 0$, for example when considering individual quark contributions to the magnetic moments.

Terms describing the explicit symmetry breaking at leading order in the quark masses are generated by
\begin{align}
\label{eq:LinLag}
\nonumber
\mathcal{L}_{\textrm{lin}}=\mathcal{B}\frac{e}{2m_N}\left[ \vphantom{\left(\overline{B}^{ijk}\right)}\right.&c_1 \left(\overline{B} m_\psi B\right)\textrm{Str}(Q) + c_2\left(\overline{B}B m_\psi \right) \textrm{Str}(Q) \\ \nonumber
& + c_3 \left( \overline{B} Q B \right) \textrm{Str}(m_\psi) + c_4 \left(\overline{B} B Q \right) \textrm{Str}(m_\psi) \\ \nonumber
& + c_5 \left( \overline{B} Qm_\psi B \right)  + c_6 \left(\overline{B} B Qm_\psi \right) \\ \nonumber
& + c_7 \left(\overline{B}B \right)\textrm{Str}(Q m_\psi) + c_8 \left(\overline{B}B\right)\textrm{Str}(Q)\textrm{Str}(m_\psi) \\ \nonumber
& + c_9 (-1)^{\eta_l(\eta_j+\eta_m)}\left(\overline{B}^{kji}(m_\psi)_i^l Q_j^m B_{lmk}\right) \\ \nonumber
& + c_{10} (-1)^{\eta_j \eta_m +1} \left( \overline{B}^{kji} (m_\psi)_i^m Q_j^l B_{lmk}\right) \\ \nonumber
& + c_{11} (-1)^{\eta_l(\eta_j+\eta_m)}\left(\overline{B}^{kji}Q_i^l (m_\psi)_j^m B_{lmk}\right) \\
& +  \left.c_{12} (-1)^{\eta_j \eta_m +1} \left( \overline{B}^{kji}Q_i^m (m_\psi)_j^l B_{lmk}\right)\right]F_{\mu\nu}\sigma^{\mu\nu},
\end{align}
where $\mathcal{B}=4\lambda/f^2$ (see Eq.~(\ref{eq:lambdaeq})), the shorthand for field bilinear invariants is summarized in Appendix~\ref{app:FBI},
\begin{figure}
\begin{center}
\subfigure[]{\label{fig:mesinsoct}\includegraphics[width=0.23\textwidth]{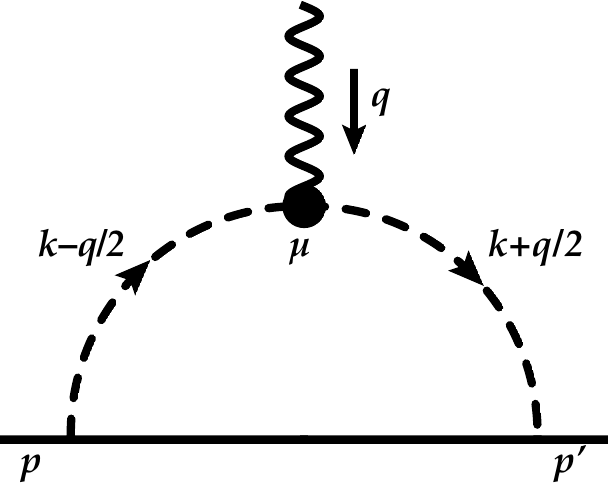}}
\subfigure[]{\label{fig:mesinsdec}\includegraphics[width=0.23\textwidth]{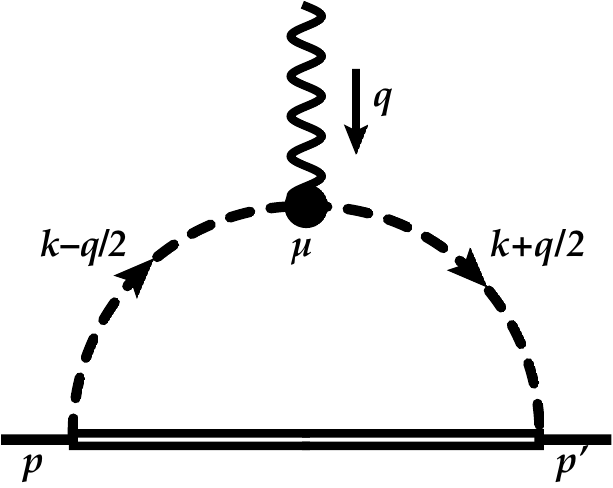}}
\caption{Loop diagrams which contribute to $G_M$ at leading order. Single, double, dashed and wavy lines represent octet baryons, decuplet baryons, mesons and photons respectively.}
\label{fig:mesinsloopst}
\end{center}
\end{figure}
and the one-loop diagrams in Fig.~\ref{fig:mesinsloopst} give rise to the leading chiral nonanalyticities of the quark mass expansion.

For small momentum transfer, the standard perturbative approach would be to generate extensions of Eqs.~\ref{eg:MagM} and \ref{eq:LinLag}, with additional derivatives, to form a series expansion in $Q^2$. In the present work we are interested in the form factors over a much larger range of $Q^2$ than can be explored with a perturbative expansion. 
For this reason we consider independent chiral extrapolations at \emph{fixed} values of $Q^2$. 

We take a model that maintains the SU(3) flavor structure of Eqs.~\ref{eg:MagM} and \ref{eq:LinLag}.
The parameters $\mu_{\alpha,\beta,\gamma}$ appearing in Eq.~\ref{eg:MagM} are now interpreted as chiral limit form factors at some fixed $Q^2$; their numerical values may be different at each $Q^2$. Similarly, the terms of Eq.~\ref{eq:LinLag} are associated with the symmetry breaking at fixed $Q^2$.

The resulting expressions for the magnetic form factors, at some fixed finite $Q^2$, may be summarized as
\begin{multline}
\label{eq:FitFunc}
G_M^{B,q}(Q^2) =  \alpha^{Bq} + \sum_{q'} \overline{\alpha}^{Bq(q')} \mathcal{B}m_{q'}  \\
 + \frac{m_N}{16 \pi^3 f^2} \sum_\phi \left(  \beta^{Bq(\phi)}_{O} \mathcal{I}_{O}(m_\phi,Q^2)+ \beta^{Bq(\phi)}_{D} \mathcal{I}_{D}(m_\phi,Q^2)\right),
\end{multline}
where $\mathcal{B}m_q$ denotes the mass of the quark $q$, identified with the meson masses via the appropriate Gell-Mann-Oakes-Renner relation, e.g., $\mathcal{B}m_l =m_\pi^2/2$. The physical mass of the nucleon is given by $m_N$ and $\phi$ stands for any of the 80 pseudo-Goldstone mesons of our theory. The pion decay constant is $f=0.0871$ GeV in the chiral limit~\cite{Amoros200187}. We note that this expression is defined in units of physical nuclear magnetons $\mu_N$.
Here the contributions from Figs. \ref{fig:mesinsoct} and \ref{fig:mesinsdec} depend on the integrals
\begin{align}
\label{eq:IO}
I_O  &=  \int d\vec{k} \frac{k_y^2 u(\vec{k}+\vec{q}/2)u(\vec{k}-\vec{q}/2)}{ 2\omega_+^2 \omega_-^2} \\ \label{eq:ID}
I_D & = \int d\vec{k} \frac{k_y^2(\omega_-+\omega_++\delta)u(\vec{k}+\vec{q}/2)u(\vec{k}-\vec{q}/2)}{2(\omega_++\delta)(\omega_-+\delta)\omega_+\omega_-(\omega_++\omega_-)},
\end{align}
where
\begin{align}
\omega_{\pm}=\sqrt{(\vec{k}\pm\vec{q}/2)^2+m^2}
\end{align}
and $u(\vec{k})$ is the ultra-violet regulator used in the finite-range regularization (FRR) scheme. This choice of regularization procedure is discussed in detail in Refs.~\cite{Leinweber:2003dg,Young:2002cj,Young:2002ib}. In short, the inclusion of a finite cutoff into the loop integrands effectively resums the chiral expansion in a way that suppresses the loop contributions at large meson masses. This enforces the physical expectation, based on the finite size of the baryon, that meson emission and absorption processes are suppressed for large momenta. For the case of the octet baryon masses, FRR appears to offer markedly improved convergence properties of the (traditionally poorly convergent) SU(3) chiral expansion~\cite{Leinweber:2003dg}, and this scheme consistently provides robust fits to lattice data at leading or next-to-leading order. Nevertheless, one could calculate the size of higher order corrections to confirm that these contributions are small as expected.

For this analysis we choose a dipole regulator $u(k)=\left(\frac{\Lambda^2}{\Lambda^2+k^2}\right)^2$ with a regulator mass $\Lambda=0.8\pm 0.1$~GeV. The dipole form is suggested by a comparison of the nucleon's axial and induced pseudoscalar form factors~\cite{Guichon:1982zk} and the choice of $\Lambda$ is informed by a lattice analysis of nucleon magnetic moments~\cite{Hall:2012pk}. 
We note that different regulator forms, for example monopole, Gaussian or sharp cutoff yield fit parameters (and extrapolated results) which are consistent within the quoted uncertainties.
Expressions for the coefficients $\alpha^{Bq}$, $\overline{\alpha}^{Bq(q')}$, $\beta_O^{Bq(\phi)}$ and $\beta_D^{Bq(\phi)}$ are given explicitly in Appendix~\ref{app:ExtrapDetails}. 

\section{Fits to lattice results}
\label{sec:Fits}

Before fitting chiral expressions to the lattice simulation results, we perform several corrections to the raw lattice data. First, we estimate corrections for small finite-volume effects, using the leading one-loop results of the chiral EFT (see Sec.~\ref{sec:FVCorrections}). As the chiral extrapolation functions summarized in Sec.~\ref{sec:ChiPTextrap} are for fixed, finite $Q^2$, we analyze the lattice results in fixed $Q^2$ bins. 
As explained in Sec.~\ref{sec:Q2bins} below, there are small variations in $Q^2$ with different pseudoscalar and baryon masses. In order to facilitate the fixed--$Q^2$ extrapolation, we interpolate the form factors to common points in $Q^2$.

All of the analysis is performed for the magnetic form factors $G_M$ in physical nuclear magnetons. This choice simplifies the extrapolation procedure as there is no need to consider a quark-mass dependent magneton although an extrapolation using such units is possible and equivalent. The conversion from lattice natural magnetons to physical nuclear magnetons is performed at the bootstrap level.

\subsection{Finite-volume corrections}
\label{sec:FVCorrections}

Finite-volume corrections are performed using the difference between infinite-volume integrals and finite-volume sums for the leading-order loop integral expressions from Sec.~\ref{sec:ChiPTextrap}. The procedure used here follows Ref.~\cite{Hall:2013oga}. We note that before performing the finite-volume sums, the expressions for the integrands in Eqs.~(\ref{eq:IO}, \ref{eq:ID}) are shifted from being symmetric (meson lines with momenta $k-q/2$ and $k+q/2$, as illustrated in Fig.~\ref{fig:mesinsloopst}) to what is more natural for the lattice, namely meson lines with momenta $k$ and $k+q$. The purpose is to account for the fact that momentum is quantized on the lattice.

The finite-volume corrections are small: they contribute approximately $2-4\%$ of the nucleon form factor at the lowest $Q^2$ value ($\approx 0.26~$GeV$^2$) and $0.03-0.06\%$ at the largest ($Q^2\approx 1.35$~GeV$^2$), where the variation in each range is a result of the different pion and kaon mass points considered.

\subsection{Binning in $Q^2$}
\label{sec:Q2bins}

\begin{figure}
\begin{center}
\includegraphics[width=0.48\textwidth]{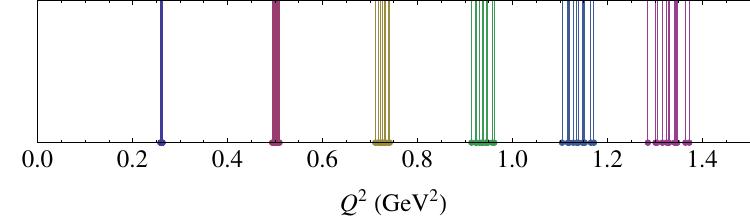}
\caption{$Q^2$ distribution for the lattice simulation results. Colors indicate the $Q^2$ bin groupings; each bin corresponds to a single value of the three-momentum transfer in lattice units.}
\label{fig:Q2BinPlots}
\end{center}
\end{figure}

As the chiral extrapolations used here (Eq.~(\ref{eq:FitFunc})) are applicable for fixed finite $Q^2$, we bin the lattice simulation results in $Q^2$ before fitting. The bin groupings are illustrated in Fig.~\ref{fig:Q2BinPlots}. Each bin corresponds to a single value of the three-momentum transfer in lattice units. The corresponding physical $Q^2$ values vary slightly because of the different baryon masses feeding into the dispersion relation. The largest variation is 1.29$-$1.37~GeV$^2$ for the highest $Q^2$ bin.

\begin{figure}
\begin{center}
\includegraphics[width=0.48\textwidth]{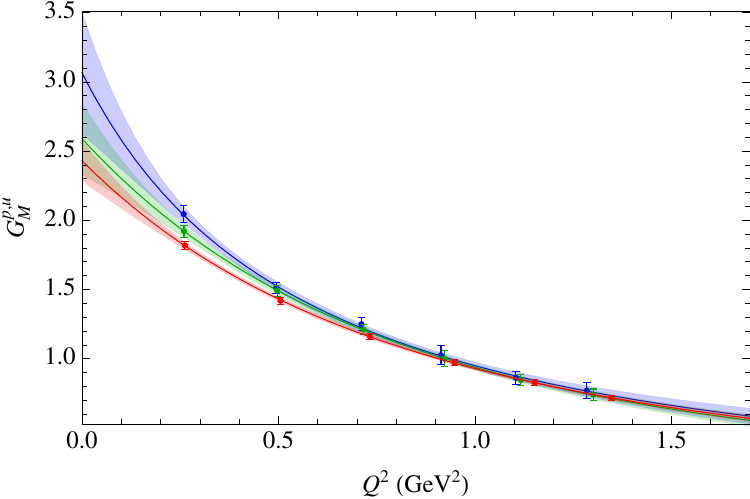}
\caption{Generalized dipole fits (Eq.~(\ref{eq:gendipfit})) upon which the binning corrections are based. The three fits shown correspond to the three different pseudoscalar mass points along the primary simulation trajectory (simulations 1--3 in Table~\ref{tab:SimDetails}). Quarks have unit charge.}
\label{fig:DipoleFits}
\end{center}
\end{figure}

To account for the small variation in $Q^2$ within each bin, all simulation results are shifted to the average $Q^2$ value of their respective bin. This shift is performed using a dipole-like fit to the (finite-volume--corrected) simulation results.
The functional form used is
\begin{equation}
\label{eq:gendipfit}
G^\textrm{fit}_M(Q^2)=\frac{\mu}{1+d_1Q^2+d_2Q^4},
\end{equation}
where $\mu$, $d_1$ and $d_2$ are free parameters. Several examples of the fits are shown in Fig.~\ref{fig:DipoleFits}. As the shifts are small, particularly at low $Q^2$ where the fit function has a larger slope, there is no dependence, within uncertainties, on the functional form chosen.
The simulation results are shifted by $G^\textrm{fit}(Q^2_\textrm{average})-G^\textrm{fit}(Q^2_\textrm{simulation})$.

\subsection{Fits to lattice results}

\begin{figure}
\begin{center}
\includegraphics[width=0.48\textwidth]{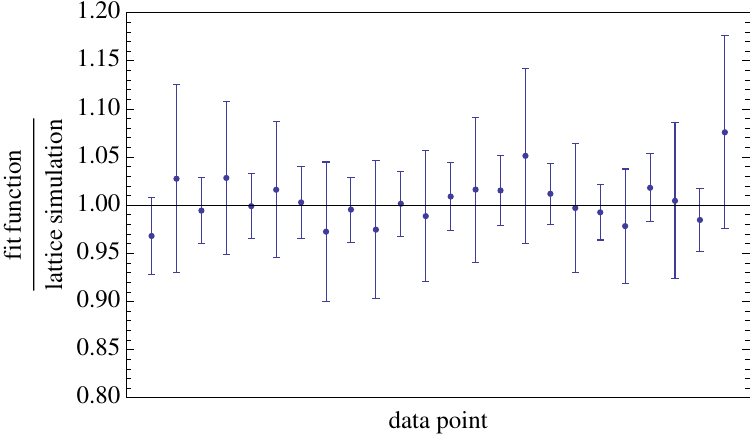}
\caption{Illustration of the quality fit for the data set at $Q^2 \approx 0.26$~GeV$^2$, the lowest $Q^2$ bin. Each point denotes one of the lattice simulation results e.g., $G_M^{p,u}$, $G_M^{p,d}$ \ldots, at one of the sets of pseudoscalar masses.}
\label{fig:FitQual}
\end{center}
\end{figure}

After the lattice simulation results have been finite-volume corrected and binned in $Q^2$, we perform an independent bootstrap-level fit, using Eq.~(\ref{eq:FitFunc}), to the variation with $m_\pi$ for the results in each $Q^2$ bin. 
An advantage of this approach~\cite{Wang:2008vb,Wang:2007iw} is that it allows the fit parameters, which are the undetermined chiral coefficients, to vary with $Q^2$ without the need to impose some phenomenological expectation on the shape of their variation. Values of these fit parameters are shown in Appendix~\ref{app:fitParams}.
The quality of fit at each $Q^2$ is good, with $\chi^2/\textrm{d.o.f.} \approx 0.5-1.4$ for every fit. An illustration of the fit quality for the lowest $Q^2$ bin ($Q^2\approx 0.26$~GeV$^2$) is given in Fig.~\ref{fig:FitQual}. That figure shows the ratio of the fit function to the lattice simulation result for each data point; the 24 data points include 6 at each set of pseudoscalar masses where $m_\pi\ne m_K$ (i.e., $G_M^{p,u}$, $G_M^{p,d}$, $G_M^{\Sigma,u}$, $G_M^{\Sigma,s}$, $G_M^{\Xi,s}$ and $G_M^{\Xi,u}$) and 2 at each SU(3)-symmetric point. We recall that while each $Q^2$ set is treated as independent, the various octet baryon form factors 
are fit simultaneously. 

\begin{figure}[]
\begin{center}
\includegraphics[width=0.48\textwidth]{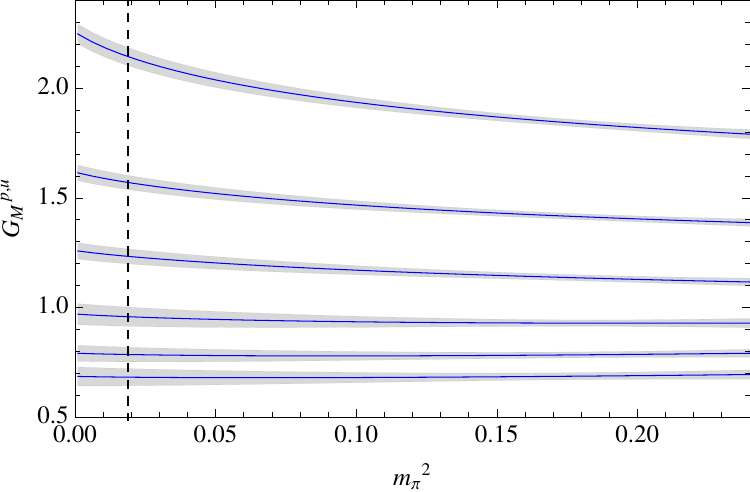}
\caption{Up quark (connected) contribution to the proton magnetic form factor for quarks with unit charge. Each set of results (top to bottom) represents the fit at a different (increasing) $Q^2$ value. The lines show these fits evaluated along the trajectory which holds the singlet pseudoscalar mass $(m_K^2+m_\pi^2/2)$ fixed to its physical value. 
}
\label{fig:TotalPuPlot}
\end{center}
\end{figure}

Using these fits, the baryon magnetic form factors may be extrapolated to the physical pseudoscalar masses at each simulation $Q^2$.
For example, Fig.~\ref{fig:TotalPuPlot} shows results for the up quark contribution to the proton magnetic form factor, %
plotted along a trajectory which holds the singlet pseudoscalar mass ($m_K^2+m_\pi^2/2$) fixed to its physical value.
The results display the expected qualitative behavior; as $Q^2$ increases (moving down the figure), the extrapolation in $m_\pi^2$ decreases in curvature. 
This implies that the magnetic radius of the proton increases in magnitude as we approach the physical pion mass from above. Magnetic radii are discussed further in Sec.~\ref{sec:radii}.

We note that uncertainty in the value of the lattice scale $a$ affects the values of both the form factors and $Q^2$ in physical units. At low $Q^2$ the shift in the form factors, and at high $Q^2$ the shift in $Q^2$ itself, is not negligible when varying $a=0.074(2)$ within the quoted uncertainties.  
Nevertheless, repeating the analysis presented in the following sections for $a$ values at the extremities of the quoted range yields fits which are almost indistinguishable from those presented for the central value -- essentially the points are shifted a short distance along the $Q^2$ fit lines -- and give entirely consistent results for each quantity, even when extrapolated to $Q^2=0$.

\section{Analysis of results}
\label{sec:Results}

In this section we summarize the results of the chiral extrapolations. In particular, we focus on isovector quantities which do not suffer from corrections associated with disconnected quark loops (section~\ref{sec:isovector}), connected octet baryon magnetic moments (Sec.~\ref{sec:magmoments}) and magnetic radii (Sec.~\ref{sec:radii}). 
Comparison of the results with experimental determinations of these quantities gives some insight into the size of disconnected contributions to the magnetic form factors.

\subsection{Isovector quantities}
\label{sec:isovector}

Isovector quantities are of particular interest as they have the advantage that contributions from disconnected quark loops, omitted in the lattice simulations, cancel. It is therefore these isovector quantities which we can determine with the smallest systematic uncertainty.

\begin{figure}[t]
\begin{center}
\includegraphics[width=0.48\textwidth]{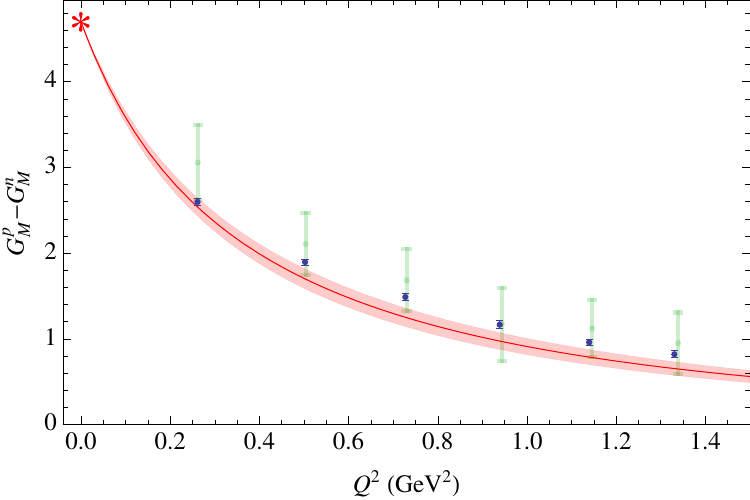}
\caption{Isovector nucleon magnetic form factor compared to the Kelly parameterization of experimental results~\cite{Kelly:2004hm}. The small (solid  blue) points show the results including all lattice simulations, while the large error bars (pale green) show the results including only lattice simulations along the primary simulation trajectory (see Table~\ref{tab:SimDetails}).
}
\label{fig:KellyIsovec}
\end{center}
\end{figure}

The agreement of the extrapolated isovector baryon form factors with experimental results is impressive.
In particular, Fig.~\ref{fig:KellyIsovec} compares the isovector nucleon form factor extracted from this analysis with the experimental determination as parameterised by Kelly~\cite{Kelly:2004hm}. While there is a tendency for the extrapolated values to be slightly high overall, the agreement, across the entire range of $Q^2$ values considered, is remarkable. We note that the uncertainties shown for the Kelly parameterization may be overestimated as we were unable to take into account the effect of correlations between the fit parameters.

\begin{figure}[t]
\begin{center}
\subfigure[Isovector sigma baryon magnetic form factor.]{
\includegraphics[width=0.48\textwidth]{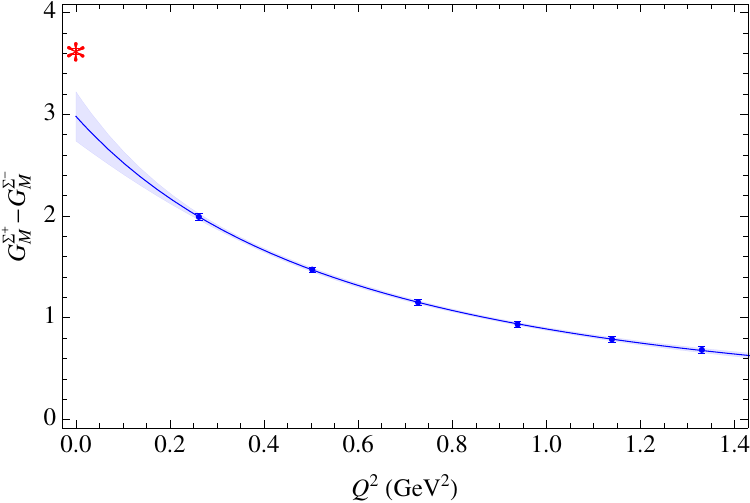}
\label{fig:IsoSig}
}
\subfigure[Isovector cascade baryon magnetic form factor.]{
\includegraphics[width=0.48\textwidth]{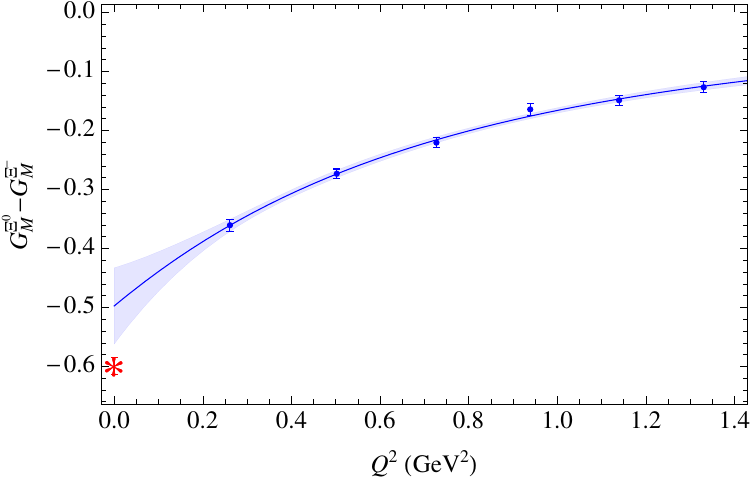}
\label{fig:IsoXi}
}
\caption{Isovector sigma and cascade baryon magnetic form factor with dipole-like fits (Eq.~(\ref{eq:gendipfit})). The red stars indicate the experimental isovector magnetic moments.}
\label{fig:IsoSigXi}
\end{center}
\end{figure}

The isovector combinations of sigma and cascade baryon magnetic form factors are shown in Figs.~\ref{fig:IsoSig} and \ref{fig:IsoXi}. As no experimental results are available for these form factors apart from $Q^2=0$, dipole-like fits (Eq.~(\ref{eq:gendipfit})) to the extrapolated simulation results, as well as the experimental isovector baryon magnetic moments, are shown. Again we find fair agreement with the experimentally measured baryon magnetic moments at $Q^2=0$, even with simple phenomenological fits parameterizing the $Q^2$-dependence of the form factors. It is clear, however, that slightly greater curvature in the $Q^2$ fit functions would improve the agreement with experiment. Isovector magnetic moments, extracted using these fits, are given in Table~\ref{tab:MagMomentsIso}.

We emphasize that lattice simulation results away from the primary simulation trajectory (see Fig.~\ref{fig:mlmsPlot}) are essential to tightly constrain the chiral extrapolations to the physical point. The effect of adding the additional off-trajectory points to the fit -- a factor of $\approx 6$ reduction in statistical uncertainty -- is shown in Fig.~\ref{fig:KellyIsovec}. This illustrates the importance for chiral extrapolations of performing lattice simulations which map out the $m_l-m_s$ plane, rather than simply following a single trajectory in this space.

\begin{table}[]
\caption{\label{tab:MagMomentsIso} Extrapolated results for the isovector magnetic moments, based on the fit to the lattice simulation results. A dipole-like parameterization (Eq.~(\ref{eq:gendipfit})) has been used for the $Q^2$-dependence. }
\begin{ruledtabular}
\begin{tabular}{lddd}
 & \multicolumn{3}{c}{$\mu_B$~($\mu_N$)}\\
\multicolumn{1}{c}{$B$} & \multicolumn{1}{c}{$p-n$} & \multicolumn{1}{c}{$\Sigma^+-\Sigma^-$} & \multicolumn{1}{c}{$\Xi^0-\Xi^-$} \\
\hline
Extrapolated & 3.8(3) & 3.0(2) & -0.51(8) \\
Experimental & 4.706 & 3.62(3) & -0.60(1) \\
\end{tabular}
\end{ruledtabular}
\end{table}

\subsection{Connected quantities}
\label{sec:magmoments}

As well as the isovector quantities presented in the previous section, we can determine the `connected part' of all individual baryon form factors. Comparison of these quantities with experimental determinations is of particular interest -- significant disconnected contributions to the form factors would cause a systematic discrepancy between the lattice and experimental results.

\begin{figure}[]
\begin{center}
\subfigure[Proton magnetic form factor.]{
\label{fig:pKelly}
\includegraphics[width=0.48\textwidth]{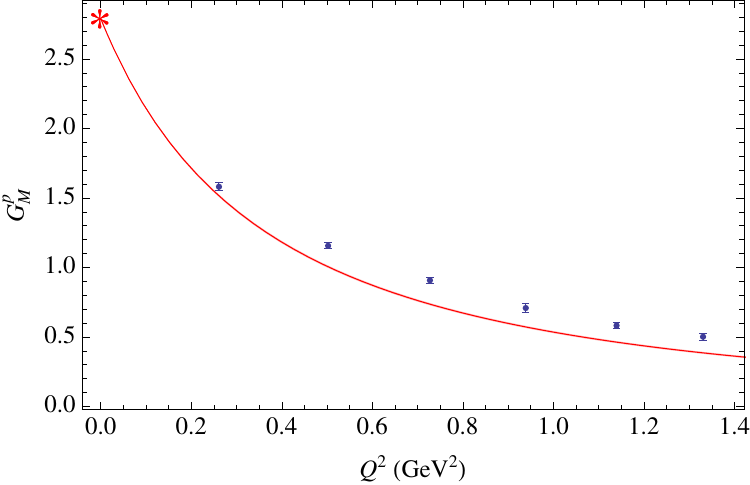}
}
\subfigure[Neutron magnetic form factor.]{
\label{fig:nKelly}
\includegraphics[width=0.48\textwidth]{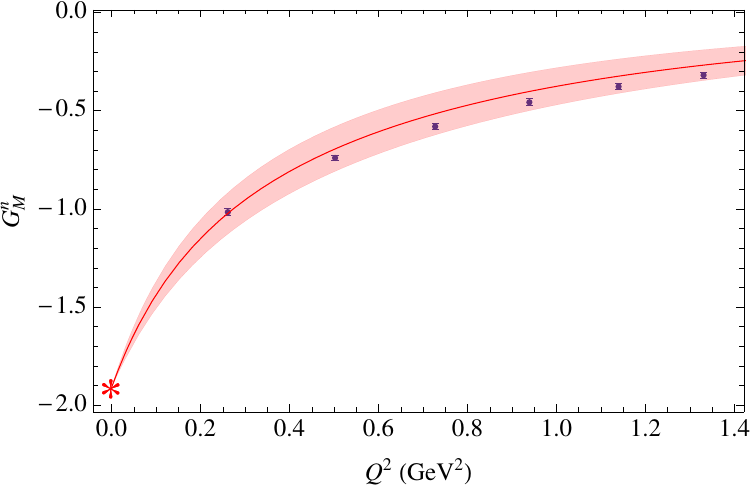}
}
\caption{Extrapolated (connected part of the) proton and neutron magnetic form factors, compared with Kelly parameterization~\cite{Kelly:2004hm} of experimental measurements.}
\end{center}
\end{figure}

Figures~\ref{fig:pKelly} and \ref{fig:nKelly} show extrapolated results for the connected parts of the proton and neutron magnetic form factors, compared with the Kelly experimental parameterization~\cite{Kelly:2004hm}. The level of agreement between the lattice and experimental results across the entire range of $Q^2$ values supports the conclusion of Ref.~\cite{Abdel-Rehim:2013wlz} that the omitted disconnected contributions are relatively small.

Figures displaying connected form factors for each of the octet baryons, including dipole-like fits in $Q^2$, are given in Appendix~\ref{app:FFPics}. The magnetic moments extracted from these fits, given in Table~\ref{tab:MagMoments}, are close to the experimental values, although we note once again that greater curvature in the $Q^2$ functional form would improve agreement with experiment. 

\begin{table*}[btf]
\caption{\label{tab:MagMoments} Results for the connected contribution to the octet baryon magnetic moments, based on a dipole-like fit (Eq.~(\ref{eq:gendipfit})) to the extrapolated lattice simulation results, compared with experimental values. }
\begin{ruledtabular}
\begin{tabular}{ldddddd}
& \multicolumn{6}{c}{$\mu_B$~($\mu_N$)}\\
\multicolumn{1}{c}{$B$} & \multicolumn{1}{c}{$p$} &  \multicolumn{1}{c}{$n$} &  \multicolumn{1}{c}{$\Sigma^+$} &  \multicolumn{1}{c}{$\Sigma^-$} &  \multicolumn{1}{c}{$\Xi^0$} &  \multicolumn{1}{c}{$\Xi^-$} \\
\hline
Extrapolated & 2.3(3) & -1.45(17) & 2.12(18) & -0.85(10) & -1.07(7) & -0.57(5) \\
Experimental & 2.79 & -1.913 & 2.458(10) & -1.160(25) & -1.250(14) & -0.6507(25) \\
\end{tabular}
\end{ruledtabular}
\end{table*}

\subsection{Magnetic radii}
\label{sec:radii}

The magnetic radii of the octet baryons are defined by
\begin{equation}
\langle r^2_M \rangle^B = -\frac{6}{G_M^B(0)}\frac{d}{dQ^2}G_M^B(Q^2)\bigg|_{Q^2=0}. 
\end{equation}
To evaluate this expression from the lattice simulation results, we use the dipole-like fits (Eq.~(\ref{eq:gendipfit})) shown in Appendix~\ref{app:FFPics}. Results, compared with available experimental data, are given in Table~\ref{tab:MagRadii}.

\begin{table*}[btf]
\caption{\label{tab:MagRadii} Extrapolated results for the octet baryon magnetic radii, based on our fit to the lattice simulation results, compared with experimental values. Results labelled `free $\mu_B$' result from a dipole-like fit function in $Q^2$ (Eq.~(\ref{eq:gendipfit})), while those labelled `general' come from the ansatz given in Eq.~(\ref{eq:GenFit}) with fixed $\mu_B$. }
\begin{ruledtabular}
\begin{tabular}{lD{.}{.}{5}D{.}{.}{4}D{.}{.}{4}D{.}{.}{4}D{.}{.}{4}D{.}{.}{4}}
 & \multicolumn{6}{c}{$\langle r^2_M \rangle^B$~(fm$^2$)}\\
 & \multicolumn{1}{c}{$p$} &  \multicolumn{1}{c}{$n$} &  \multicolumn{1}{c}{$\Sigma^+$} &  \multicolumn{1}{c}{$\Sigma^-$} &  \multicolumn{1}{c}{$\Xi^0$} &  \multicolumn{1}{c}{$\Xi^-$} \\
\hline
Extrapolated (free $\mu_B$) & 0.35(11) & 0.35(11) & 0.39(9) & 0.42(13) & 0.27(8) & 0.23(8) \\
Extrapolated (general)& 0.71(8) &  0.86(9) & 0.66(5) & 1.05(9) & 0.53(5) & 0.44(5)\\
Experimental & 0.777(16) & 0.862(9) &  &  &  &  \\
\end{tabular}
\end{ruledtabular}
\end{table*}

It is notable that we find consistently smaller values for the magnetic radii than those determined experimentally (for the nucleon) or predicted in chiral quark models (for the octet baryons)~\cite{Liu:2013fda,Silva2005290}.
This is perhaps not unexpected; comparing Figs.~\ref{fig:pKelly} and \ref{fig:nKelly} with Figs.~\ref{fig:MMp} and \ref{fig:MMn} shows that although our results are quite consistent with the experimental parameterization of the nucleon form factors {\em where they are calculated}, the best-fit dipole function has slightly less curvature. As noted in the previous sections, greater curvature in the $Q^2$ fit forms would improve consistency with the experimental magnetic moments for all of the octet baryons. 

\begin{figure}
\begin{center}
\includegraphics[width=0.48\textwidth]{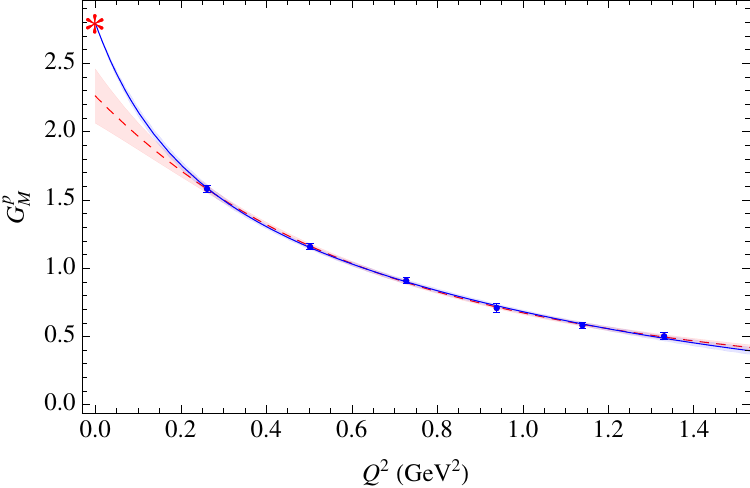}
\caption{Dipole-like (from Eq.~(\ref{eq:gendipfit}), dashed red band) and general (from Eq.~(\ref{eq:GenFit}), solid blue band) fits to the proton magnetic form factor. The quality of fit is comparable for both fits.}
\label{fig:GPComp}
\end{center}
\end{figure}

To improve the extraction of the magnetic radii, we consider a second functional form in $Q^2$, inspired by the Kelly-style parameterizations of experimental results with a more general polynomial in the denominator:
\begin{equation}
G^B_M(Q^2)=\frac{\mu_B}{1+c Q^2 + d Q^4 + f Q^6}.
\label{eq:GenFit}
\end{equation}
We now fix $\mu_B$ to the experimental magnetic moment, so there are again three free parameters, $c$, $d$ and $f$. As illustrated for the proton in Fig.~\ref{fig:GPComp}, the quality of fit using this functional form is entirely comparable with that for the dipole-like fit. The shift in the extracted value of the magnetic radius, however, is significant, as shown in Table~\ref{tab:MagRadii}. This example confirms that truly robust predictions for the hyperon magnetic radii from lattice QCD will require results at much lower $Q^2$ values to eliminate the significant dependence on the functional form chosen for the $Q^2$ extrapolation. 
Nevertheless, the level of agreement of the extracted nucleon magnetic radii with experimental values indicates that, by taking the experimental magnetic moments as additional input, we have achieved the first accurate calculation of the magnetic radii of the entire outer ring of the baryon octet from lattice QCD, extrapolated to the physical pseudoscalar masses. 

\subsection{Quark form factors}

The chiral extrapolations discussed in previous sections are in fact performed for the individual doubly and singly represented quark contributions to the magnetic form factors. Inspecting these contributions can give insight into the environmental sensitivity of the distribution of quarks inside a hadron.

\begin{figure}[]
\begin{center}
\subfigure[Doubly-represented quark contributions.]{
\includegraphics[width=0.48\textwidth]{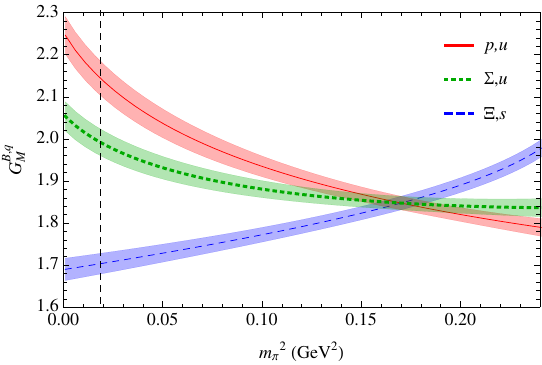}
\label{fig:DoublyRep}
}
\subfigure[Singly-represented quark contributions.]{
\label{fig:SinglyRep}
\includegraphics[width=0.48\textwidth]{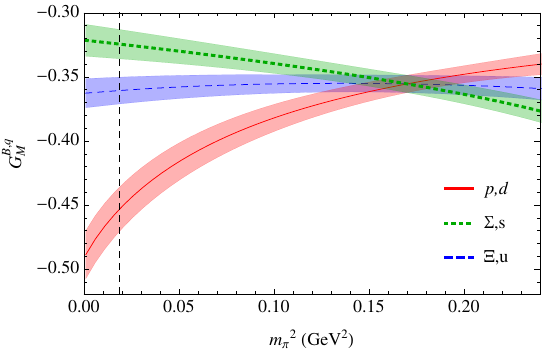}
}
\caption{Connected part of the doubly and singly-represented quark contributions to the baryon magnetic form factors for $Q^2\approx 0.26$~GeV$^2$. The charges of the relevant quarks have been set to unity.}
\end{center}
\end{figure}

Chiral extrapolations for the connected part of these quark contributions, shown along the trajectory which holds the singlet pseudoscalar mass $(m_K^2+m_\pi^2/2)$ fixed to its physical value, are presented in Figs.~\ref{fig:DoublyRep} and \ref{fig:SinglyRep}. The figures show the lowest $Q^2$ result, at approximately 0.26~GeV$^2$. Of course, the fits shown are simultaneously constrained by the lattice simulation results for all of the octet baryons at that $Q^2$.
 
Comparison of the $u$ quark contributions to the proton and $\Sigma^+$ in Fig.~\ref{fig:DoublyRep} shows the relative suppression of $G_M^{\Sigma,u}$ caused by the heavier spectator quark in the sigma. This effect is replicated, and is more significant, when probing the singly represented quark, as can be seen by the relative suppression (in magnitude) of the $u$ contribution to the cascade baryon compared to the $d$ in the proton in Fig.~\ref{fig:SinglyRep}. Changing the mass of the probed quark -- doubly represented in the proton compared with the cascade, or singly represented in the proton compared with the sigma -- causes a similar effect.

\section{Conclusion}

We have presented the results of a 2+1-flavor lattice QCD study of the electromagnetic form factors of the octet baryons. Calculations are performed on one volume with a single lattice spacing, six different sets of pseudoscalar masses and six values of $Q^2$ in the range 0.2--1.3~GeV$^2$. The Dirac and Pauli radii of the nucleon, extracted using generalized dipole fits, are in line with other recent 2+1 and 2+1+1 flavor lattice calculations with similar values of the pion mass.

By performing lattice simulations on configurations which `map out' the $m_l-m_s$ plane, rather than following a single trajectory in this space, we are able to robustly constrain a chiral extrapolation of the magnetic Sachs form factor $G_M$ to the physical pseudoscalar masses at each simulation $Q^2$.
Systematic uncertainties are controlled by performing finite-volume corrections. The uncertainties inherent in the determination of the lattice scale $a$, the shape of the ultra-violet cutoff and the value of the cutoff parameter $\Lambda$ in the finite-range regularization scheme are found to be negligible. 

As such, the single most significant limitation of this calculation is that disconnected quark loops are omitted from the lattice simulations. For this reason isovector combinations, where contributions from disconnected quark loops cancel, are of significant interest.
The nucleon isovector form factor extracted from this work compares well with the experimental results over the entire range of $Q^2$ values considered. 
It is notable that the precision of these results is such that it is foreseeable that this generation of lattice QCD simulations will rival experiment in terms of precision.

The proton and neutron magnetic form factors from this work, which include only the `connected' quark loop contributions, agree rather well with the experimental determinations at all simulation $Q^2$ values. The comparison with experiment is also favourable for the magnetic moments and magnetic radii of the rest of the outer-ring baryon octet, extracted using a dipole-like form in $Q^2$.
This suggests that the omitted disconnected quark loop contributions are small relative to the uncertainties of this calculation.

We point out that a pure dipole form in $Q^2$ does not, in general, provide a good fit to the lattice simulation results. A dipole-like function with a more general polynomial in the denominator is significantly better, as described above. A comparison of nucleon observables extracted using both fit forms indicates that the dipole yields significantly poorer predictions for the magnetic moments and radii, despite the form factors matching the experimental values at larger $Q^2$.
This suggests that meaningful extractions of the magnetic moments and radii from lattice QCD require a more careful analysis than the standard procedure using a dipole fit in $Q^2$, unless simulations are performed for very small $Q^2$ values much less than 0.2~GeV$^2$. Analyses similar to that performed here may reveal that other existing lattice simulations are in fact more compatible with experiment than the results of the standard calculations indicate.

\section*{Acknowledgements}

The numerical configuration generation was performed using the BQCD lattice QCD program~\cite{Nakamura:2010qh} on the IBM BlueGeneQ using DIRAC 2 resources (EPCC, Edinburgh, UK), the BlueGene P and Q at NIC (J\"ulich, Germany) and the SGI ICE 8200 at HLRN (Berlin-Hannover, Germany). The BlueGene codes were optimised using Bagel~\cite{Boyle:2009vp}. The Chroma software library~\cite{Edwards:2004sx} was used in the data analysis. This work was supported by the EU grants 283286 (HadronPhysics3), 227431 (Hadron Physics2) and by the University of Adelaide and the Australian
Research Council through the ARC Centre of Excellence for Particle Physics at the Terascale and grants FL0992247 (AWT), DP110101265 (RDY), FT120100821 (RDY) and FT100100005 (JMZ).

\appendix
\cleardoublepage
\begin{widetext}
\section{Lattice simulation results}
\label{app:rawLatticeResults}

This section presents tables of lattice simulation results for $F_1$ and $F_2$ for the simulation parameters described in Sec.~\ref{sec:LatticeSimulation}. 
Results for the Dirac and Pauli mean-squared charge radii $\langle r^2 \rangle_{1,2}^{B,q}$ and anomalous magnetic moments $\kappa^{B,q}$, discussed in Sec.~\ref{subsec:F1F2res}, are shown in Tables~\ref{tab:r1}--\ref{tab:kappa}.

\begin{table*}[!Hb]
\begin{ruledtabular}
\begin{tabular}{cccccccc}
$m_\pi$ (MeV) & $m_K$ (MeV) & $\langle r^2 \rangle_1^{p,u}$~(fm$^2$) & $\langle r^2 \rangle_1^{p,d}$ & $\langle r^2 \rangle_1^{\Sigma,u}$ & $\langle r^2 \rangle_1^{\Sigma,s}$ & $\langle r^2 \rangle_1^{\Xi,s}$ &  $\langle r^2 \rangle_1^{\Xi,u}$ \\ \hline
$465$ & $465$ & $\text{0.334 (16)}$ & $\text{0.387 (22)}$ & $\text{0.334 (16)}$ & $\text{0.387 (22)}$ & $\text{0.334 (16)}$ & $\text{0.387 (22)}$ \\
$360$ & $505$ & $\text{0.368 (11)}$ & $\text{0.420 (12)}$ & $\text{0.3639 (87)}$ & $\text{0.3630 (60)}$ & $\text{0.3218 (58)}$ & $\text{0.4260 (84)}$ \\
$310$ & $520$ & $\text{0.376 (20)}$ & $\text{0.437 (24)}$ & $\text{0.399 (13)}$ & $\text{0.382 (10)}$ & $\text{0.3329 (65)}$ & $\text{0.459 (11)}$ \\
$440$ & $440$ & $\text{0.3601 (96)}$ & $\text{0.405 (12)}$ & $\text{0.3601 (96)}$ & $\text{0.405 (12)}$ & $\text{0.3601 (96)}$ & $\text{0.405 (12)}$ \\
$400$ & $400$ & $\text{0.378 (10)}$ & $\text{0.438 (15)}$ & $\text{0.378 (10)}$ & $\text{0.438 (15)}$ & $\text{0.378 (10)}$ & $\text{0.438 (15)}$ \\
$330$ & $435$ & $\text{0.396 (13)}$ & $\text{0.445 (25)}$ & $\text{0.400 (10)}$ & $\text{0.412 (14)}$ & $\text{0.3650 (74)}$ & $\text{0.465 (13)}$ \\
\end{tabular}
\end{ruledtabular}
\caption{Dirac mean-squared charge radii, extracted from generalized dipole fits -- see Sec.~\ref{subsec:F1F2res}.}
\label{tab:r1}
\end{table*}

\begin{table*}[!Hb]
\begin{ruledtabular}
\begin{tabular}{cccccccc}
$m_\pi$ (MeV) & $m_K$ (MeV) & $\langle r^2 \rangle_2^{p,u}$~(fm$^2$) & $\langle r^2 \rangle_2^{p,d}$ & $\langle r^2 \rangle_2^{\Sigma,u}$ & $\langle r^2 \rangle_2^{\Sigma,s}$ & $\langle r^2 \rangle_2^{\Xi,s}$ &  $\langle r^2 \rangle_2^{\Xi,u}$ \\ \hline
$465$ & $465$ & $\text{0.337 (18)}$ & $\text{0.3434 (79)}$ & $\text{0.337 (18)}$ & $\text{0.3434 (79)}$ & $\text{0.337 (18)}$ & $\text{0.3434 (79)}$ \\
$360$ & $505$ & $\text{0.335 (29)}$ & $\text{0.405 (18)}$ & $\text{0.340 (20)}$ & $\text{0.3358 (99)}$ & $\text{0.292 (15)}$ & $\text{0.389 (10)}$ \\
$310$ & $520$ & $\text{0.364 (59)}$ & $\text{0.379 (32)}$ & $\text{0.331 (29)}$ & $\text{0.286 (14)}$ & $\text{0.282 (15)}$ & $\text{0.367 (14)}$ \\
$440$ & $440$ & $\text{0.491 (51)}$ & $\text{0.415 (22)}$ & $\text{0.491 (51)}$ & $\text{0.415 (22)}$ & $\text{0.491 (51)}$ & $\text{0.415 (22)}$ \\
$400$ & $400$ & $\text{0.377 (48)}$ & $\text{0.362 (26)}$ & $\text{0.377 (48)}$ & $\text{0.362 (26)}$ & $\text{0.377 (48)}$ & $\text{0.362 (26)}$ \\
$330$ & $435$ & $\text{0.429 (51)}$ & $\text{0.416 (28)}$ & $\text{0.413 (37)}$ & $\text{0.361 (17)}$ & $\text{0.369 (26)}$ & $\text{0.387 (14)}$ \\  
\end{tabular}
\end{ruledtabular}
\caption{Pauli mean-squared charge radii, extracted from generalized dipole fits -- see Sec.~\ref{subsec:F1F2res}.}
\label{tab:r2}
\end{table*}

\begin{table*}[!Hb]
\begin{ruledtabular}
\begin{tabular}{cccccccc}
$m_\pi$ (MeV) & $m_K$ (MeV) & $\kappa^{p,u}$~($\mu_N$) & $\kappa^{p,d}$ & $\kappa^{\Sigma,u}$ & $\kappa^{\Sigma,s}$ & $\kappa^{\Xi,s}$ &  $\kappa^{\Xi,u}$ \\ \hline
$465$ & $465$ & $\text{1.331 (43)}$ & $\text{-1.570 (24)}$ & $\text{1.331 (43)}$ & $\text{-1.570 (24)}$ & $\text{1.331 (43)}$ & $\text{-1.570 (24)}$ \\
$360$ & $505$ & $\text{1.171 (62)}$ & $\text{-1.624 (46)}$ & $\text{1.421 (54)}$ & $\text{-1.590 (25)}$ & $\text{1.284 (36)}$ & $\text{-1.675 (28)}$ \\
$310$ & $520$ & $\text{1.24 (13)}$ & $\text{-1.526 (80)}$ & $\text{1.480 (82)}$ & $\text{-1.469 (33)}$ & $\text{1.305 (38)}$ & $\text{-1.616 (36)}$ \\
$440$ & $440$ & $\text{1.35 (10)}$ & $\text{-1.652 (62)}$ & $\text{1.35 (10)}$ & $\text{-1.652 (62)}$ & $\text{1.35 (10)}$ & $\text{-1.652 (62)}$ \\
$400$ & $400$ & $\text{1.29 (11)}$ & $\text{-1.464 (67)}$ & $\text{1.29 (11)}$ & $\text{-1.464 (67)}$ & $\text{1.29 (11)}$ & $\text{-1.464 (67)}$ \\
$330$ & $435$ & $\text{1.32 (11)}$ & $\text{-1.582 (69)}$ & $\text{1.439 (93)}$ & $\text{-1.557 (41)}$ & $\text{1.320 (62)}$ & $\text{-1.598 (36)}$ \\
\end{tabular}
\end{ruledtabular}
\caption{Anomalous magnetic moments in nuclear magnetons, extracted from generalized dipole fits -- see Sec.~\ref{subsec:F1F2res}.}
\label{tab:kappa}
\end{table*}

\end{widetext}

\begin{table*}
\begin{ruledtabular}
\begin{tabular}{ccD{.}{.}{-1}D{.}{.}{1}D{.}{.}{1}D{.}{.}{1}D{.}{.}{1}}
$m_\pi$ (MeV) & $m_K$ (MeV) & \multicolumn{1}{c}{$Q^2$ (GeV$^2$)} & \multicolumn{1}{c}{$F_1^{p,u}$} & \multicolumn{1}{c}{$F_1^{p,d}$} & \multicolumn{1}{c}{$F_2^{p,u}$} & \multicolumn{1}{c}{$F_2^{p,d}$} \\ \hline
$465$ & $465$ & $0.26$ & $\text{1.434 (24)}$ & $\text{0.666 (11)}$ & $\text{0.932 (20)}$ & $\text{-1.113 (11)}$ \\
$\text{}$ & $\text{}$ & $0.51$ & $\text{1.134 (19)}$ & $\text{0.4873 (94)}$ & $\text{0.722 (18)}$ & $\text{-0.8298 (94)}$ \\
$\text{}$ & $\text{}$ & $0.73$ & $\text{0.936 (17)}$ & $\text{0.3744 (88)}$ & $\text{0.589 (19)}$ & $\text{-0.6525 (88)}$ \\
$\text{}$ & $\text{}$ & $0.95$ & $\text{0.804 (16)}$ & $\text{0.3014 (75)}$ & $\text{0.474 (21)}$ & $\text{-0.5547 (75)}$ \\
$\text{}$ & $\text{}$ & $1.15$ & $\text{0.697 (15)}$ & $\text{0.2491 (72)}$ & $\text{0.392 (16)}$ & $\text{-0.4621 (72)}$ \\
$\text{}$ & $\text{}$ & $1.35$ & $\text{0.616 (15)}$ & $\text{0.2058 (73)}$ & $\text{0.328 (15)}$ & $\text{-0.3956 (73)}$ \\
$360$ & $505$ & $0.26$ & $\text{1.3982 (91)}$ & $\text{0.6425 (40)}$ & $\text{0.822 (28)}$ & $\text{-1.081 (18)}$ \\
$\text{}$ & $\text{}$ & $0.51$ & $\text{1.089 (12)}$ & $\text{0.4588 (51)}$ & $\text{0.651 (23)}$ & $\text{-0.792 (12)}$ \\
$\text{}$ & $\text{}$ & $0.72$ & $\text{0.884 (17)}$ & $\text{0.3412 (66)}$ & $\text{0.535 (26)}$ & $\text{-0.622 (13)}$ \\
$\text{}$ & $\text{}$ & $0.92$ & $\text{0.781 (32)}$ & $\text{0.284 (11)}$ & $\text{0.396 (36)}$ & $\text{-0.527 (24)}$ \\
$\text{}$ & $\text{}$ & $1.12$ & $\text{0.656 (26)}$ & $\text{0.2219 (81)}$ & $\text{0.341 (22)}$ & $\text{-0.426 (17)}$ \\
$\text{}$ & $\text{}$ & $1.3$ & $\text{0.551 (26)}$ & $\text{0.1719 (81)}$ & $\text{0.324 (23)}$ & $\text{-0.339 (15)}$ \\
$310$ & $520$ & $0.26$ & $\text{1.382 (18)}$ & $\text{0.6253 (75)}$ & $\text{0.885 (58)}$ & $\text{-1.034 (33)}$ \\
$\text{}$ & $\text{}$ & $0.49$ & $\text{1.075 (20)}$ & $\text{0.4433 (82)}$ & $\text{0.620 (39)}$ & $\text{-0.792 (24)}$ \\
$\text{}$ & $\text{}$ & $0.71$ & $\text{0.883 (29)}$ & $\text{0.316 (13)}$ & $\text{0.528 (41)}$ & $\text{-0.586 (34)}$ \\
$\text{}$ & $\text{}$ & $0.91$ & $\text{0.754 (41)}$ & $\text{0.268 (15)}$ & $\text{0.409 (59)}$ & $\text{-0.519 (38)}$ \\
$\text{}$ & $\text{}$ & $1.1$ & $\text{0.633 (29)}$ & $\text{0.194 (11)}$ & $\text{0.346 (34)}$ & $\text{-0.435 (25)}$ \\
$\text{}$ & $\text{}$ & $1.29$ & $\text{0.535 (36)}$ & $\text{0.158 (17)}$ & $\text{0.343 (43)}$ & $\text{-0.342 (30)}$ \\
$440$ & $440$ & $0.26$ & $\text{1.3994 (79)}$ & $\text{0.6540 (40)}$ & $\text{0.823 (38)}$ & $\text{-1.080 (24)}$ \\
$\text{}$ & $\text{}$ & $0.5$ & $\text{1.078 (11)}$ & $\text{0.4689 (56)}$ & $\text{0.590 (31)}$ & $\text{-0.804 (20)}$ \\
$\text{}$ & $\text{}$ & $0.73$ & $\text{0.871 (15)}$ & $\text{0.3548 (79)}$ & $\text{0.451 (31)}$ & $\text{-0.623 (21)}$ \\
$\text{}$ & $\text{}$ & $0.94$ & $\text{0.733 (21)}$ & $\text{0.2827 (92)}$ & $\text{0.336 (32)}$ & $\text{-0.479 (20)}$ \\
$\text{}$ & $\text{}$ & $1.14$ & $\text{0.616 (19)}$ & $\text{0.2264 (89)}$ & $\text{0.270 (24)}$ & $\text{-0.403 (17)}$ \\
$\text{}$ & $\text{}$ & $1.33$ & $\text{0.545 (25)}$ & $\text{0.189 (11)}$ & $\text{0.236 (23)}$ & $\text{-0.349 (20)}$ \\
$400$ & $400$ & $0.26$ & $\text{1.3974 (91)}$ & $\text{0.6411 (53)}$ & $\text{0.854 (56)}$ & $\text{-1.027 (29)}$ \\
$\text{}$ & $\text{}$ & $0.5$ & $\text{1.084 (12)}$ & $\text{0.4564 (62)}$ & $\text{0.692 (38)}$ & $\text{-0.744 (24)}$ \\
$\text{}$ & $\text{}$ & $0.72$ & $\text{0.888 (20)}$ & $\text{0.3377 (89)}$ & $\text{0.506 (33)}$ & $\text{-0.596 (25)}$ \\
$\text{}$ & $\text{}$ & $0.93$ & $\text{0.787 (28)}$ & $\text{0.286 (12)}$ & $\text{0.412 (47)}$ & $\text{-0.533 (28)}$ \\
$\text{}$ & $\text{}$ & $1.13$ & $\text{0.668 (20)}$ & $\text{0.2299 (85)}$ & $\text{0.361 (32)}$ & $\text{-0.411 (21)}$ \\
$\text{}$ & $\text{}$ & $1.32$ & $\text{0.585 (27)}$ & $\text{0.184 (10)}$ & $\text{0.296 (26)}$ & $\text{-0.356 (26)}$ \\
$330$ & $435$ & $0.26$ & $\text{1.367 (11)}$ & $\text{0.6303 (80)}$ & $\text{0.819 (46)}$ & $\text{-1.029 (28)}$ \\
$\text{}$ & $\text{}$ & $0.5$ & $\text{1.057 (14)}$ & $\text{0.437 (10)}$ & $\text{0.651 (30)}$ & $\text{-0.773 (16)}$ \\
$\text{}$ & $\text{}$ & $0.72$ & $\text{0.875 (17)}$ & $\text{0.324 (13)}$ & $\text{0.511 (31)}$ & $\text{-0.593 (20)}$ \\
$\text{}$ & $\text{}$ & $0.92$ & $\text{0.726 (33)}$ & $\text{0.267 (16)}$ & $\text{0.340 (45)}$ & $\text{-0.473 (31)}$ \\
$\text{}$ & $\text{}$ & $1.12$ & $\text{0.614 (26)}$ & $\text{0.207 (13)}$ & $\text{0.296 (27)}$ & $\text{-0.395 (21)}$ \\
$\text{}$ & $\text{}$ & $1.3$ & $\text{0.544 (29)}$ & $\text{0.170 (13)}$ & $\text{0.271 (30)}$ & $\text{-0.319 (24)}$ 
\end{tabular}
\end{ruledtabular}
\caption{Raw lattice simulation results for the nucleon.}
\end{table*}

\begin{table*}
\begin{ruledtabular}
\begin{tabular}{ccD{.}{.}{-1}D{.}{.}{1}D{.}{.}{1}D{.}{.}{1}D{.}{.}{1}}
$m_\pi$ (MeV) & $m_K$ (MeV) & \multicolumn{1}{c}{$Q^2$ (GeV$^2$)} & \multicolumn{1}{c}{$F_1^{\Sigma,u}$} & \multicolumn{1}{c}{$F_1^{\Sigma,s}$} & \multicolumn{1}{c}{$F_2^{\Sigma,u}$} & \multicolumn{1}{c}{$F_2^{\Sigma,s}$} \\ \hline
$465$ & $465$ & $0.26$ & $\text{1.434 (24)}$ & $\text{0.666 (11)}$ & $\text{0.932 (20)}$ & $\text{-1.113 (11)}$ \\
$\text{}$ & $\text{}$ & $0.51$ & $\text{1.134 (19)}$ & $\text{0.4873 (94)}$ & $\text{0.722 (18)}$ & $\text{-0.8298 (94)}$ \\
$\text{}$ & $\text{}$ & $0.73$ & $\text{0.936 (17)}$ & $\text{0.3744 (88)}$ & $\text{0.589 (19)}$ & $\text{-0.6525 (88)}$ \\
$\text{}$ & $\text{}$ & $0.95$ & $\text{0.804 (16)}$ & $\text{0.3014 (75)}$ & $\text{0.474 (21)}$ & $\text{-0.5547 (75)}$ \\
$\text{}$ & $\text{}$ & $1.15$ & $\text{0.697 (15)}$ & $\text{0.2491 (72)}$ & $\text{0.392 (16)}$ & $\text{-0.4621 (72)}$ \\
$\text{}$ & $\text{}$ & $1.35$ & $\text{0.616 (15)}$ & $\text{0.2058 (73)}$ & $\text{0.328 (15)}$ & $\text{-0.3956 (73)}$ \\
$360$ & $505$ & $0.26$ & $\text{1.4008 (72)}$ & $\text{0.6829 (21)}$ & $\text{0.996 (24)}$ & $\text{-1.126 (10)}$ \\
$\text{}$ & $\text{}$ & $0.5$ & $\text{1.0839 (97)}$ & $\text{0.5058 (31)}$ & $\text{0.770 (21)}$ & $\text{-0.8620 (89)}$ \\
$\text{}$ & $\text{}$ & $0.73$ & $\text{0.871 (13)}$ & $\text{0.3882 (43)}$ & $\text{0.615 (20)}$ & $\text{-0.680 (10)}$ \\
$\text{}$ & $\text{}$ & $0.95$ & $\text{0.774 (23)}$ & $\text{0.3301 (73)}$ & $\text{0.479 (27)}$ & $\text{-0.587 (15)}$ \\
$\text{}$ & $\text{}$ & $1.15$ & $\text{0.646 (20)}$ & $\text{0.2611 (60)}$ & $\text{0.414 (19)}$ & $\text{-0.479 (13)}$ \\
$\text{}$ & $\text{}$ & $1.34$ & $\text{0.545 (21)}$ & $\text{0.2092 (68)}$ & $\text{0.367 (18)}$ & $\text{-0.393 (13)}$ \\
$310$ & $520$ & $0.26$ & $\text{1.372 (12)}$ & $\text{0.6776 (36)}$ & $\text{1.062 (38)}$ & $\text{-1.095 (14)}$ \\
$\text{}$ & $\text{}$ & $0.51$ & $\text{1.055 (14)}$ & $\text{0.5074 (56)}$ & $\text{0.796 (25)}$ & $\text{-0.855 (17)}$ \\
$\text{}$ & $\text{}$ & $0.73$ & $\text{0.855 (20)}$ & $\text{0.3937 (82)}$ & $\text{0.657 (29)}$ & $\text{-0.681 (24)}$ \\
$\text{}$ & $\text{}$ & $0.95$ & $\text{0.731 (24)}$ & $\text{0.327 (10)}$ & $\text{0.507 (35)}$ & $\text{-0.592 (21)}$ \\
$\text{}$ & $\text{}$ & $1.15$ & $\text{0.641 (22)}$ & $\text{0.2667 (94)}$ & $\text{0.439 (25)}$ & $\text{-0.515 (20)}$ \\
$\text{}$ & $\text{}$ & $1.35$ & $\text{0.563 (30)}$ & $\text{0.222 (14)}$ & $\text{0.419 (33)}$ & $\text{-0.442 (27)}$ \\
$440$ & $440$ & $0.26$ & $\text{1.3994 (79)}$ & $\text{0.6540 (40)}$ & $\text{0.823 (38)}$ & $\text{-1.080 (24)}$ \\
$\text{}$ & $\text{}$ & $0.5$ & $\text{1.078 (11)}$ & $\text{0.4689 (56)}$ & $\text{0.590 (31)}$ & $\text{-0.804 (20)}$ \\
$\text{}$ & $\text{}$ & $0.73$ & $\text{0.871 (15)}$ & $\text{0.3548 (79)}$ & $\text{0.451 (31)}$ & $\text{-0.623 (21)}$ \\
$\text{}$ & $\text{}$ & $0.94$ & $\text{0.733 (21)}$ & $\text{0.2827 (92)}$ & $\text{0.336 (32)}$ & $\text{-0.479 (20)}$ \\
$\text{}$ & $\text{}$ & $1.14$ & $\text{0.616 (19)}$ & $\text{0.2264 (89)}$ & $\text{0.270 (24)}$ & $\text{-0.403 (17)}$ \\
$\text{}$ & $\text{}$ & $1.33$ & $\text{0.545 (25)}$ & $\text{0.189 (11)}$ & $\text{0.236 (23)}$ & $\text{-0.349 (20)}$ \\
$400$ & $400$ & $0.26$ & $\text{1.3974 (91)}$ & $\text{0.6411 (53)}$ & $\text{0.854 (56)}$ & $\text{-1.027 (29)}$ \\
$\text{}$ & $\text{}$ & $0.5$ & $\text{1.084 (12)}$ & $\text{0.4564 (62)}$ & $\text{0.692 (38)}$ & $\text{-0.744 (24)}$ \\
$\text{}$ & $\text{}$ & $0.72$ & $\text{0.888 (20)}$ & $\text{0.3377 (89)}$ & $\text{0.506 (33)}$ & $\text{-0.596 (25)}$ \\
$\text{}$ & $\text{}$ & $0.93$ & $\text{0.787 (28)}$ & $\text{0.286 (12)}$ & $\text{0.412 (47)}$ & $\text{-0.533 (28)}$ \\
$\text{}$ & $\text{}$ & $1.13$ & $\text{0.668 (20)}$ & $\text{0.2299 (85)}$ & $\text{0.361 (32)}$ & $\text{-0.411 (21)}$ \\
$\text{}$ & $\text{}$ & $1.32$ & $\text{0.585 (27)}$ & $\text{0.184 (10)}$ & $\text{0.296 (26)}$ & $\text{-0.356 (26)}$ \\
$330$ & $435$ & $0.26$ & $\text{1.3678 (86)}$ & $\text{0.6557 (48)}$ & $\text{0.915 (41)}$ & $\text{-1.076 (16)}$ \\
$\text{}$ & $\text{}$ & $0.5$ & $\text{1.053 (11)}$ & $\text{0.4731 (66)}$ & $\text{0.714 (24)}$ & $\text{-0.815 (13)}$ \\
$\text{}$ & $\text{}$ & $0.73$ & $\text{0.864 (13)}$ & $\text{0.3598 (81)}$ & $\text{0.555 (27)}$ & $\text{-0.633 (17)}$ \\
$\text{}$ & $\text{}$ & $0.94$ & $\text{0.734 (24)}$ & $\text{0.297 (11)}$ & $\text{0.414 (34)}$ & $\text{-0.529 (20)}$ \\
$\text{}$ & $\text{}$ & $1.14$ & $\text{0.624 (22)}$ & $\text{0.238 (10)}$ & $\text{0.343 (23)}$ & $\text{-0.442 (17)}$ \\
$\text{}$ & $\text{}$ & $1.33$ & $\text{0.554 (27)}$ & $\text{0.198 (11)}$ & $\text{0.296 (24)}$ & $\text{-0.368 (21)}$ 
\end{tabular}
\end{ruledtabular}
\caption{Raw lattice simulation results for the sigma baryon.}
\end{table*}

\begin{table*}
\begin{ruledtabular}
\begin{tabular}{ccD{.}{.}{-1}D{.}{.}{1}D{.}{.}{1}D{.}{.}{1}D{.}{.}{1}}
$m_\pi$ (MeV) & $m_K$ (MeV) & \multicolumn{1}{c}{$Q^2$ (GeV$^2$)} & \multicolumn{1}{c}{$F_1^{\Xi,s}$} & \multicolumn{1}{c}{$F_1^{\Xi,u}$} & \multicolumn{1}{c}{$F_2^{\Xi,s}$} & \multicolumn{1}{c}{$F_2^{\Xi,u}$} \\ \hline
$465$ & $465$ & $0.26$ & $\text{1.434 (24)}$ & $\text{0.666 (11)}$ & $\text{0.932 (20)}$ & $\text{-1.113 (11)}$ \\
$\text{}$ & $\text{}$ & $0.51$ & $\text{1.134 (19)}$ & $\text{0.4873 (94)}$ & $\text{0.722 (18)}$ & $\text{-0.8298 (94)}$ \\
$\text{}$ & $\text{}$ & $0.73$ & $\text{0.936 (17)}$ & $\text{0.3744 (88)}$ & $\text{0.589 (19)}$ & $\text{-0.6525 (88)}$ \\
$\text{}$ & $\text{}$ & $0.95$ & $\text{0.804 (16)}$ & $\text{0.3014 (75)}$ & $\text{0.474 (21)}$ & $\text{-0.5547 (75)}$ \\
$\text{}$ & $\text{}$ & $1.15$ & $\text{0.697 (15)}$ & $\text{0.2491 (72)}$ & $\text{0.392 (16)}$ & $\text{-0.4621 (72)}$ \\
$\text{}$ & $\text{}$ & $1.35$ & $\text{0.616 (15)}$ & $\text{0.2058 (73)}$ & $\text{0.328 (15)}$ & $\text{-0.3956 (73)}$ \\
$360$ & $505$ & $0.26$ & $\text{1.4537 (51)}$ & $\text{0.6457 (27)}$ & $\text{0.940 (18)}$ & $\text{-1.129 (10)}$ \\
$\text{}$ & $\text{}$ & $0.51$ & $\text{1.1536 (76)}$ & $\text{0.4607 (35)}$ & $\text{0.747 (15)}$ & $\text{-0.8270 (78)}$ \\
$\text{}$ & $\text{}$ & $0.74$ & $\text{0.948 (10)}$ & $\text{0.3437 (45)}$ & $\text{0.616 (14)}$ & $\text{-0.6411 (82)}$ \\
$\text{}$ & $\text{}$ & $0.96$ & $\text{0.841 (20)}$ & $\text{0.2909 (69)}$ & $\text{0.481 (18)}$ & $\text{-0.531 (13)}$ \\
$\text{}$ & $\text{}$ & $1.17$ & $\text{0.712 (19)}$ & $\text{0.2278 (58)}$ & $\text{0.422 (16)}$ & $\text{-0.436 (11)}$ \\
$\text{}$ & $\text{}$ & $1.36$ & $\text{0.608 (20)}$ & $\text{0.1789 (65)}$ & $\text{0.376 (16)}$ & $\text{-0.354 (11)}$ \\
$310$ & $520$ & $0.26$ & $\text{1.4475 (58)}$ & $\text{0.6317 (38)}$ & $\text{0.974 (18)}$ & $\text{-1.114 (13)}$ \\
$\text{}$ & $\text{}$ & $0.51$ & $\text{1.1557 (86)}$ & $\text{0.4468 (51)}$ & $\text{0.762 (16)}$ & $\text{-0.825 (11)}$ \\
$\text{}$ & $\text{}$ & $0.74$ & $\text{0.960 (13)}$ & $\text{0.3347 (78)}$ & $\text{0.630 (18)}$ & $\text{-0.640 (14)}$ \\
$\text{}$ & $\text{}$ & $0.96$ & $\text{0.834 (17)}$ & $\text{0.2742 (66)}$ & $\text{0.513 (18)}$ & $\text{-0.524 (16)}$ \\
$\text{}$ & $\text{}$ & $1.17$ & $\text{0.728 (18)}$ & $\text{0.2169 (61)}$ & $\text{0.442 (17)}$ & $\text{-0.449 (17)}$ \\
$\text{}$ & $\text{}$ & $1.37$ & $\text{0.647 (26)}$ & $\text{0.179 (10)}$ & $\text{0.403 (21)}$ & $\text{-0.376 (20)}$ \\
$440$ & $440$ & $0.26$ & $\text{1.3994 (79)}$ & $\text{0.6540 (40)}$ & $\text{0.823 (38)}$ & $\text{-1.080 (24)}$ \\
$\text{}$ & $\text{}$ & $0.5$ & $\text{1.078 (11)}$ & $\text{0.4689 (56)}$ & $\text{0.590 (31)}$ & $\text{-0.804 (20)}$ \\
$\text{}$ & $\text{}$ & $0.73$ & $\text{0.871 (15)}$ & $\text{0.3548 (79)}$ & $\text{0.451 (31)}$ & $\text{-0.623 (21)}$ \\
$\text{}$ & $\text{}$ & $0.94$ & $\text{0.733 (21)}$ & $\text{0.2827 (92)}$ & $\text{0.336 (32)}$ & $\text{-0.479 (20)}$ \\
$\text{}$ & $\text{}$ & $1.14$ & $\text{0.616 (19)}$ & $\text{0.2264 (89)}$ & $\text{0.270 (24)}$ & $\text{-0.403 (17)}$ \\
$\text{}$ & $\text{}$ & $1.33$ & $\text{0.545 (25)}$ & $\text{0.189 (11)}$ & $\text{0.236 (23)}$ & $\text{-0.349 (20)}$ \\
$400$ & $400$ & $0.26$ & $\text{1.3974 (91)}$ & $\text{0.6411 (53)}$ & $\text{0.854 (56)}$ & $\text{-1.027 (29)}$ \\
$\text{}$ & $\text{}$ & $0.5$ & $\text{1.084 (12)}$ & $\text{0.4564 (62)}$ & $\text{0.692 (38)}$ & $\text{-0.744 (24)}$ \\
$\text{}$ & $\text{}$ & $0.72$ & $\text{0.888 (20)}$ & $\text{0.3377 (89)}$ & $\text{0.506 (33)}$ & $\text{-0.596 (25)}$ \\
$\text{}$ & $\text{}$ & $0.93$ & $\text{0.787 (28)}$ & $\text{0.286 (12)}$ & $\text{0.412 (47)}$ & $\text{-0.533 (28)}$ \\
$\text{}$ & $\text{}$ & $1.13$ & $\text{0.668 (20)}$ & $\text{0.2299 (85)}$ & $\text{0.361 (32)}$ & $\text{-0.411 (21)}$ \\
$\text{}$ & $\text{}$ & $1.32$ & $\text{0.585 (27)}$ & $\text{0.184 (10)}$ & $\text{0.296 (26)}$ & $\text{-0.356 (26)}$ \\
$330$ & $435$ & $0.26$ & $\text{1.4094 (62)}$ & $\text{0.6283 (41)}$ & $\text{0.892 (28)}$ & $\text{-1.082 (14)}$ \\
$\text{}$ & $\text{}$ & $0.5$ & $\text{1.1030 (87)}$ & $\text{0.4418 (56)}$ & $\text{0.684 (18)}$ & $\text{-0.795 (11)}$ \\
$\text{}$ & $\text{}$ & $0.73$ & $\text{0.911 (11)}$ & $\text{0.3313 (63)}$ & $\text{0.546 (19)}$ & $\text{-0.623 (13)}$ \\
$\text{}$ & $\text{}$ & $0.95$ & $\text{0.792 (19)}$ & $\text{0.273 (10)}$ & $\text{0.430 (25)}$ & $\text{-0.501 (16)}$ \\
$\text{}$ & $\text{}$ & $1.15$ & $\text{0.677 (19)}$ & $\text{0.2178 (86)}$ & $\text{0.352 (18)}$ & $\text{-0.424 (13)}$ \\
$\text{}$ & $\text{}$ & $1.34$ & $\text{0.594 (23)}$ & $\text{0.1794 (86)}$ & $\text{0.306 (20)}$ & $\text{-0.354 (17)}$ 
\end{tabular}
\end{ruledtabular}
\caption{Raw lattice simulation results for the cascade baryon.}
\end{table*}

\FloatBarrier

\section{Field bilinear invariants}
\label{app:FBI}

We summarize here a compact notation for the field bilinear invariants, originally employed by Labrenz and Sharpe in Ref.~\cite{Labrenz:1996jy}. In the following expressions, $A$ is an operator with the transformation properties of the axial current $A_\mu$, while $\Gamma$ is an arbitrary Dirac matrix, for example the spin operator $S_\mu$.
\begin{align}
\left( \overline{B} \Gamma B \right) & \equiv \overline{B}_{kji}^\alpha\Gamma_\alpha^\beta B_{ijk,\beta}\\
\left(\overline{B}\Gamma A B \right) & \equiv \overline{B}_{kji}^\alpha\Gamma_\alpha^\beta A_{ii'}B_{i'jk,\beta}\\
\left( \overline{B}\Gamma B A \right) & \equiv \overline{B}_{kji}^\alpha\Gamma_\alpha^\beta A_{kk'} B_{ijk',\beta}\times (-1)^{(i+j)(k+k')}\\
\left( \overline{B} \Gamma A^\mu T_\mu \right) & \equiv \overline{B}_{kji}^\alpha\Gamma_\alpha^\beta A_{ii'}^\mu T_{\mu,i'jk}^\beta \\
\left(\overline{T}^\mu \Gamma T_\mu \right) & \equiv \overline{T}^\mu_{kji,\alpha}\Gamma^\alpha_\beta T^\beta_{\mu,ijk} \\
\left(\overline{T}^\mu \Gamma A^\nu T_\mu \right) & \equiv \overline{T}^\mu_{kji,\alpha}\Gamma^\alpha_\beta A^\nu_{ii'}T^\beta_{\mu,i'jk}
\end{align}

\FloatBarrier

\section{Chiral perturbation theory extrapolations}
\label{app:ExtrapDetails}

This section gives expressions for the chiral coefficients in Eq.~(\ref{eq:FitFunc}). The labels `doubly', `singly' and `other' indicate whether the quark $q'$ or $q$ is `doubly-represented', `singly-represented' or not at all represented in the baryon $B$.

\begin{table}[!htb]
\begin{center}
\begin{ruledtabular}
\begin{tabular}{cc}
\multicolumn{2}{c}{$\alpha^{Bq}$} \\
$\text{doubly}$ & $\text{singly}$ \\\hline
$2 \mu _F$ & $\mu _F-\mu _D$ \\
\end{tabular}
\end{ruledtabular}
\end{center}
\end{table}

\begin{table}[!htbp]
\begin{center}
\begin{ruledtabular}
\begin{tabular}{lccc}
\multicolumn{4}{c}{$\alpha^{Bq}$} \\
\backslashbox{$B$}{$q$} & $u$ & $d$ & $s$ \\\hline
$\Lambda$ & $\mu _F-\frac{2 \mu _D}{3}$ & $\mu _F-\frac{2 \mu _D}{3}$ & $\frac{\mu _D}{3}+\mu _F$ \\
$\Sigma ^0$ & $\mu _F$ & $\mu _F$ & $\mu _F-\mu _D$ \\
\end{tabular}
\end{ruledtabular}
\end{center}
\end{table}

\begin{table}[!h]
\begin{center}
\begin{ruledtabular}
\begin{tabular}{lc}
\multicolumn{2}{c}{$\overline{\alpha}^{Bq(q')}$} \\ 
\backslashbox{$m_{q'}$}{$q$} & $\text{doubly}$  \\ \hline
$m_{\text{doubly}}$ & $\frac{1}{6} (c_{10}+c_{11}+c_{12}+18 c_3+45 c_4+2 c_5+5 c_6+c_9)$  \\
$m_{\text{singly}}$ & $\frac{1}{6} (-2 c_{10}+c_{11}-2 c_{12}+18 c_3+45 c_4+4 c_9)$  \\
$m_{\text{other}}$ & $3 c_3+\frac{15 c_4}{2}$  \\ \hline
 & $\text{singly}$ \\ \hline
$m_{\text{doubly}}$ &  $\frac{1}{6} (-2 c_{10}+4 c_{11}-2 c_{12}+36 c_3+9 c_4+c_9)$ \\
$m_{\text{singly}}$ &  $\frac{1}{6} (36 c_3+9 c_4+4 c_5+c_6)$ \\
$m_{\text{other}}$ &  $\frac{3}{2} (4 c_3+c_4)$ \\
\end{tabular}
\end{ruledtabular}
\end{center}
\end{table}

\begin{table}[!htb]
\begin{center}
\begin{ruledtabular}
\begin{tabular}{lc}
& $\overline{\alpha}^{\Lambda q(q')}$ \\ 
\backslashbox{$m_{q'}$}{$q$} & $u$ \\ \hline
$m_u$ & $\frac{1}{4} (18 c_3+9 c_4+2 c_5+c_6)$ \\
$m_d$ & $\frac{1}{4} (-\text{c12}-c_{10}+c_{11}+18 c_3+9 c_4+c_9)$  \\
$m_s$ & $\frac{1}{4} (c_{11}+9 (2 c_3+c_4))$  \\ \hline
 & $d$  \\ \hline
$m_u$ &  $\frac{1}{4} (-\text{c12}-c_{10}+c_{11}+18 c_3+9 c_4+c_9)$  \\
$m_d$ &  $\frac{1}{4} (18 c_3+9 c_4+2 c_5+c_6)$ \\
$m_s$ &  $\frac{1}{4} (c_{11}+9 (2 c_3+c_4))$  \\ \hline
 & $s$ \\ \hline
$m_u$ &  $\frac{1}{4} (18 c_4+c_9)$ \\
$m_d$ &  $\frac{1}{4} (18 c_4+c_9)$ \\
$m_s$ &  $\frac{1}{2} (9 c_4+c_6)$ \\
\end{tabular}
\end{ruledtabular}
\end{center}
\end{table}

\begin{table}[!htb]
\begin{center}
\begin{ruledtabular}
\begin{tabular}{lccc}
& $\overline{\alpha}^{\Sigma^0 q(q')}$ \\ 
\backslashbox{$m_{q'}$}{$q$} & $u$  \\ \hline
$m_u$ & $\frac{1}{12} (18 c_3+45 c_4+2 c_5+5 c_6)$  \\
$m_d$ & $\frac{1}{12} (c_{10}+c_{11}+c_{12}+18 c_3+45 c_4+c_9)$  \\
$m_s$ & $\frac{1}{12} (-2 c_{10}+c_{11}-2 c_{12}+18 c_3+45 c_4+4 c_9)$  \\ \hline
 & $d$  \\ \hline
$m_u$  & $\frac{1}{12} (c_{10}+c_{11}+c_{12}+18 c_3+45 c_4+c_9)$  \\
$m_d$ & $\frac{1}{12} (18 c_3+45 c_4+2 c_5+5 c_6)$  \\
$m_s$ & $\frac{1}{12} (-2 c_{10}+c_{11}-2 c_{12}+18 c_3+45 c_4+4 c_9)$  \\ \hline
 & $s$  \\ \hline
$m_u$ &  $\frac{1}{12} (-2 c_{10}+4 c_{11}-2 c_{12}+72 c_3+18 c_4+c_9)$ \\
$m_d$ &  $\frac{1}{12} (-2 c_{10}+4 c_{11}-2 c_{12}+72 c_3+18 c_4+c_9)$ \\
$m_s$ &  $\frac{1}{6} (36 c_3+9 c_4+4 c_5+c_6)$ \\
\end{tabular}
\end{ruledtabular}
\end{center}
\end{table}


\begin{table}[!htbp]
\begin{center}
\begin{ruledtabular}
\begin{tabular}{lcc}
\multicolumn{3}{c}{$\beta_{O}^{B q(\phi)}$} \\
\backslashbox{$m_{\phi}$}{$q$} & $\text{doubly}$ & $\text{singly}$ \\ \hline
$m_{\text{doubly}}+m_{\text{singly}}$ & $4 \left(D^2+F^2\right)$ & $-\frac{2}{3} \left(D^2+6 D F-3 F^2\right)$ \\
$m_{\text{singly}}+m_{\text{other}}$ & $0$ & $2 (D-F)^2$ \\
$m_{\text{doubly}}+m_{\text{other}}$ & $\frac{4}{3} \left(D^2+3 F^2\right)$ & $0$ \\
$2m_{\text{doubly}}$ & $\frac{4}{3} \left(D^2+3 F^2\right)$ & $0$ \\
$2m_{\text{singly}}$ & $0$ & $2 (D-F)^2$ \\
\end{tabular}
\end{ruledtabular}
\end{center}
\end{table}

\begin{table}[!htbp]
\begin{center}
\begin{ruledtabular}
\begin{tabular}{lcc}
\multicolumn{3}{c}{$\beta_{D}^{B q(\phi)}$} \\
\backslashbox{$m_{\phi}$}{$q$} & $\text{doubly}$ & $\text{singly}$ \\ \hline
$m_{\text{doubly}}+m_{\text{singly}}$ & $\frac{2 C^2}{9}$ & $-\frac{5 C^2}{9}$ \\
$m_{\text{singly}}+m_{\text{other}}$ & $\text{}$ & $-\frac{2 C^2}{9}$ \\
$m_{\text{doubly}}+m_{\text{other}}$ & $-\frac{C^2}{9}$ & $\text{}$ \\
$2m_{\text{doubly}}$ & $-\frac{C^2}{9}$ & $\text{}$ \\
$2m_{\text{singly}}$ & $\text{}$ & $-\frac{2 C^2}{9}$ \\
\end{tabular}
\end{ruledtabular}
\end{center}
\end{table}

\cleardoublepage

\begin{table}[!h]
\begin{center}
\begin{ruledtabular}
\begin{tabular}{lcc}
\multicolumn{3}{c}{$\beta_{O}^{\Lambda q(\phi)}$} \\ 
\backslashbox{$m_{\phi}$}{$q$} & $u$ & $d$  \\ \hline
$m_u+m_d$ & $\frac{2}{9} \left(7 D^2-12 D F+9 F^2\right)$ & $\frac{2}{9} \left(7 D^2-12 D F+9 F^2\right)$  \\
$m_d+m_s$ & $\text{}$ & $\frac{2}{9} \left(D^2-12 D F+9 F^2\right)$ \\
$m_u+m_s$ & $\frac{2}{9} \left(D^2-12 D F+9 F^2\right)$ & $\text{}$  \\
$2m_u$ & $\frac{2}{9} \left(7 D^2-12 D F+9 F^2\right)$ & $\text{}$  \\
$2m_d$ & $\text{}$ & $\frac{2}{9} \left(7 D^2-12 D F+9 F^2\right)$  \\
$2m_s$ & $\text{}$ & $\text{}$  \\ \hline
 & \multicolumn{2}{c}{$s$} \\ \hline
$m_u+m_d$ &  \multicolumn{2}{c}{$\text{}$} \\
$m_d+m_s$ &  \multicolumn{2}{c}{$\frac{2}{9} \left(7 D^2+6 D F+9 F^2\right)$} \\
$m_u+m_s$ & \multicolumn{2}{c}{$\frac{2}{9} \left(7 D^2+6 D F+9 F^2\right)$} \\
$2m_u$ & \multicolumn{2}{c}{$\text{}$} \\
$2m_d$ &  \multicolumn{2}{c}{$\text{}$} \\
$2m_s$ &  \multicolumn{2}{c}{$\frac{2}{9} (D+3 F)^2$} \\
\end{tabular}
\end{ruledtabular}
\end{center}
\end{table}

\begin{table}[!h]
\begin{center}
\begin{ruledtabular}
\begin{tabular}{lccc}
\multicolumn{4}{c}{$\beta_{O}^{\Sigma^0 q(\phi)}$} \\ 
\backslashbox{$m_{\phi}$}{$q$} & $u$ & $d$ & $s$ \\ \hline
$m_u+m_d$ & $\frac{2}{3} \left(D^2+3 F^2\right)$ & $\frac{2}{3} \left(D^2+3 F^2\right)$ & $\text{}$ \\
$m_d+m_s$ & $\text{}$ & $2 \left(D^2+F^2\right)$ & $\frac{2}{3} \left(D^2-6 D F+3 F^2\right)$ \\
$m_u+m_s$ & $2 \left(D^2+F^2\right)$ & $\text{}$ & $\frac{2}{3} \left(D^2-6 D F+3 F^2\right)$ \\
$2m_u$ & $\frac{2}{3} \left(D^2+3 F^2\right)$ & $\text{}$ & $\text{}$ \\
$2m_d$ & $\text{}$ & $\frac{2}{3} \left(D^2+3 F^2\right)$ & $\text{}$ \\
$2m_s$ & $\text{}$ & $\text{}$ & $2 (D-F)^2$ \\
\end{tabular}
\end{ruledtabular}
\end{center}
\end{table}

\begin{table}[!h]
\begin{center}
\begin{ruledtabular}
\begin{tabular}{lccc}
\multicolumn{4}{c}{$\beta_{D}^{\Lambda q(\phi)}$} \\
\backslashbox{$m_{\phi}$}{$q$} & $u$ & $d$ & $s$ \\ \hline
$m_u+m_d$ & $-\frac{C^2}{6}$ & $-\frac{C^2}{6}$ & $\text{}$ \\
$m_d+m_s$ & $\text{}$ & $-\frac{C^2}{3}$ & $\frac{C^2}{6}$ \\
$m_u+m_s$ & $-\frac{C^2}{3}$ & $\text{}$ & $\frac{C^2}{6}$ \\
$2m_u$ & $-\frac{C^2}{6}$ & $\text{}$ & $\text{}$ \\
$2m_d$ & $\text{}$ & $-\frac{C^2}{6}$ & $\text{}$ \\
$2m_s$ & $\text{}$ & $\text{}$ & $\text{}$ \\
\end{tabular}
\end{ruledtabular}
\end{center}
\end{table}

\begin{table}[!h]
\begin{center}
\begin{ruledtabular}
\begin{tabular}{lccc}
\multicolumn{4}{c}{$\beta_{D}^{\Sigma^0 q(\phi)}$} \\
\backslashbox{$m_{\phi}$}{$q$} & $u$ & $d$ & $s$ \\ \hline
$m_u+m_d$ & $-\frac{C^2}{18}$ & $-\frac{C^2}{18}$ & $\text{}$ \\
$m_d+m_s$ & $\text{}$ & $\frac{C^2}{9}$ & $-\frac{7 C^2}{18}$ \\
$m_u+m_s$ & $\frac{C^2}{9}$ & $\text{}$ & $-\frac{7 C^2}{18}$ \\
$2m_u$ & $-\frac{C^2}{18}$ & $\text{}$ & $\text{}$ \\
$2m_d$ & $\text{}$ & $-\frac{C^2}{18}$ & $\text{}$ \\
$2m_s$ & $\text{}$ & $\text{}$ & $-\frac{2 C^2}{9}$ \\
\end{tabular}
\end{ruledtabular}
\end{center}
\end{table}


\section{Fit parameters}
\label{app:fitParams}

\begin{figure}[!hp]
\begin{center}
\includegraphics[width=0.48\textwidth]{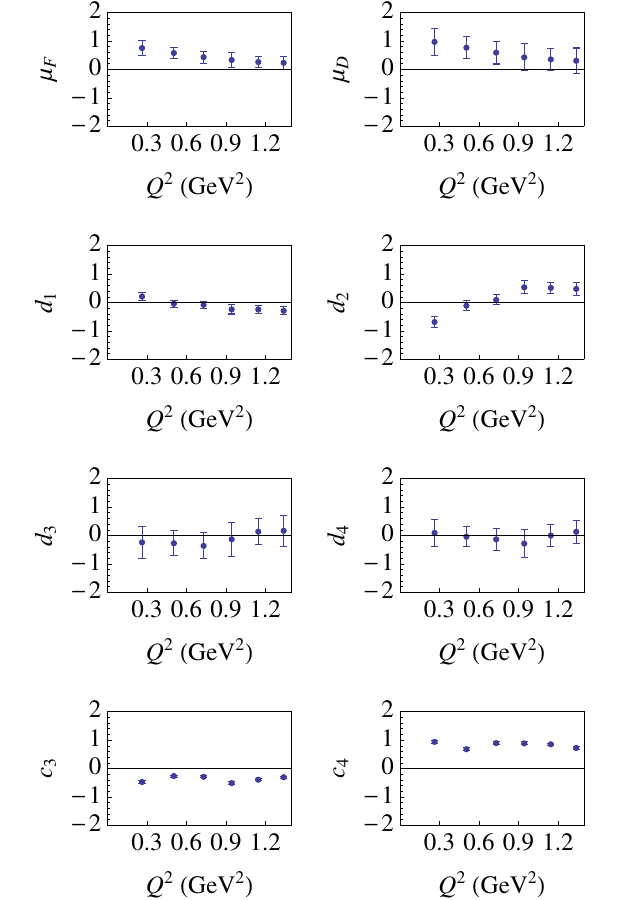}
\caption{$Q^2$ dependence of unrenormalized fit parameters, defined in Eqs.~(\ref{eq:muDF}) and (\ref{eq:LinLag}).}
\label{fig:magparams}
\end{center}
\end{figure}

Figure~\ref{fig:magparams} shows the values of the chiral parameters determined by our fits. The parameters $\mu_D$ and $\mu_F$ are defined in Eq.~(\ref{eq:muDF}), while the $c_i$ appear in Eq.~(\ref{eq:LinLag}). The $d_i$ are relevant linear combinations of the $c_i$:
\begin{align}
d_1 &=c_5-\frac{1}{4}c_{11},& d_2 &= c_6 + c_{11}, \\
d_3 &= c_6 + c_{11}, &d_4&=c_{10}-\frac{5}{2}c_4+c_{12}.
\end{align}
We note that the values of the parameters shown here are unrenormalized. They are included merely to illustrate the approximately linear $Q^2$ dependence of the parameters. Recall that the separate $Q^2$ fits are independent.

\FloatBarrier

\cleardoublepage

\begin{widetext}

\section{Octet baryon form factors - Figures}
\label{app:FFPics}

Figure~\ref{fig:octFFs} shows the connected part of the octet baryon form factors, extrapolated to the physical pseudoscalar masses. The fits shown are those used in Secs.~\ref{sec:magmoments} and \ref{sec:radii} to extract the magnetic moments and radii.


\begin{figure*}[!h]
\begin{center}
\subfigure[]{
\includegraphics[width=0.48\textwidth]{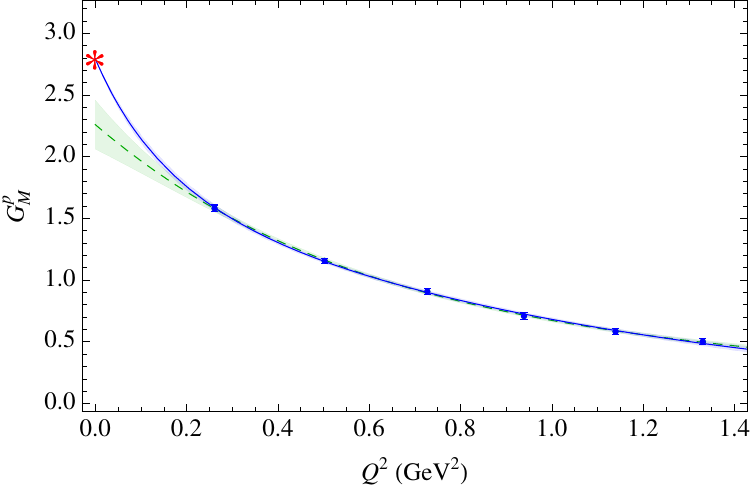}
\label{fig:MMp}
}
\subfigure[]{
\includegraphics[width=0.48\textwidth]{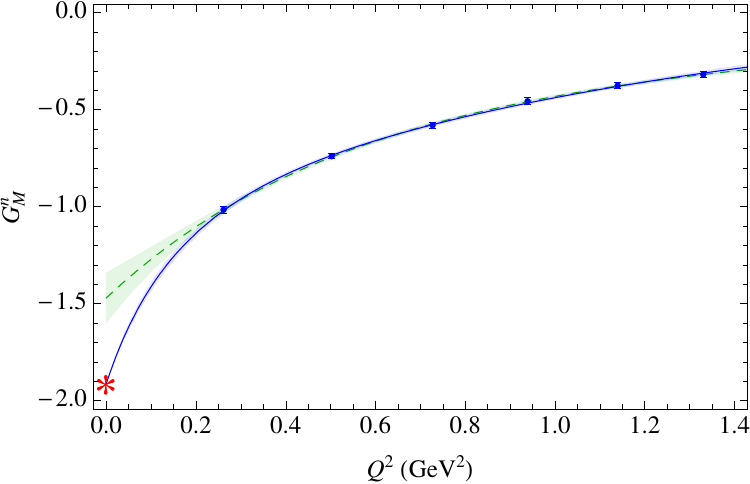}
\label{fig:MMn}
}
\subfigure[]{
\includegraphics[width=0.48\textwidth]{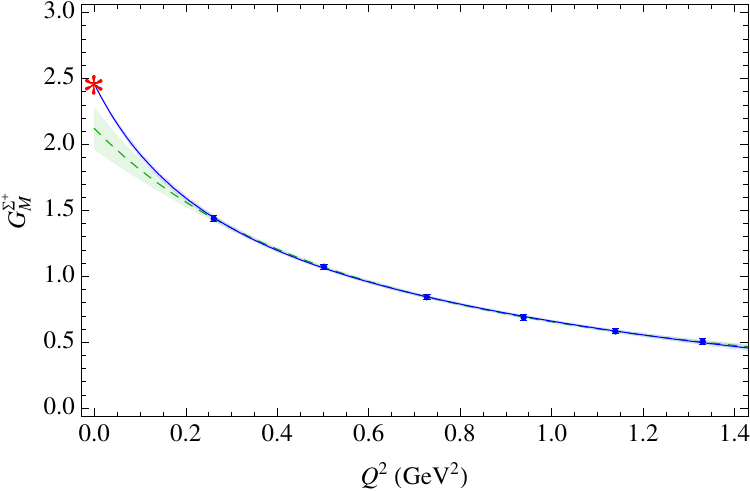}
\label{fig:MMSp}
}
\subfigure[]{
\includegraphics[width=0.48\textwidth]{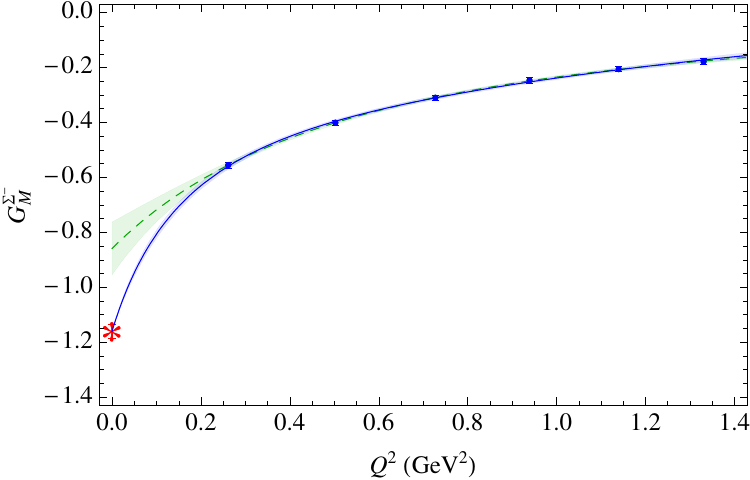}
\label{fig:MMSm}
}
\subfigure[]{
\includegraphics[width=0.48\textwidth]{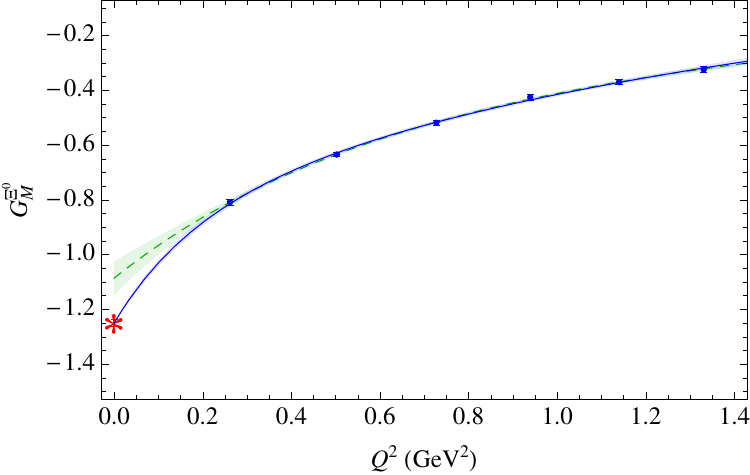}
\label{fig:MMX0}
}
\subfigure[]{
\includegraphics[width=0.48\textwidth]{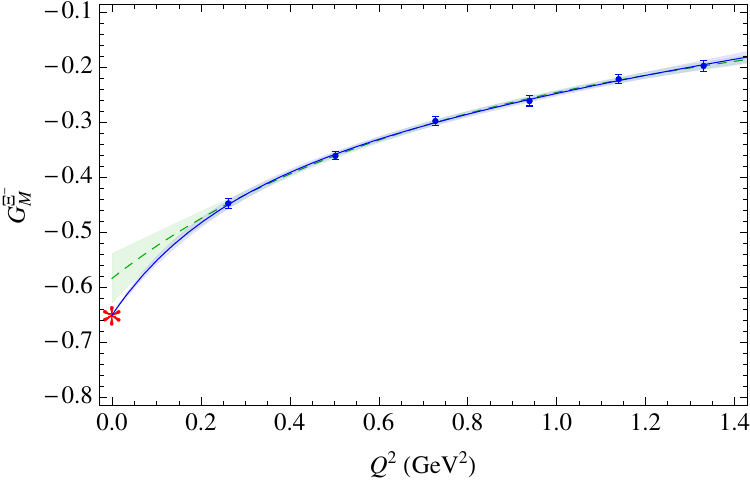}
\label{fig:MMXm}
}
\caption{Connected part of the octet baryon magnetic form factors. The red stars indicate the experimental magnetic moments. The lines show dipole-like fits (Eq.~(\ref{eq:gendipfit}), dashed green, and Eq.~(\ref{eq:GenFit}), solid blue).}
\label{fig:octFFs}
\end{center}
\end{figure*}

\end{widetext}

\cleardoublepage

\bibliography{MagFFBib}

\end{document}